\documentclass[fleqn,usenatbib]{mnras}

\usepackage[T1]{fontenc}
\usepackage{ae,aecompl}

\usepackage{graphicx}	
\usepackage{amsmath}	
\usepackage{amssymb}	
\usepackage{threeparttable}

\newcommand{\beq}	{\begin{equation}}
\newcommand{\eeq}	{\end{equation}}
\newcommand{\beqa}{\begin{eqnarray}}
\newcommand{\eeqa}{\end{eqnarray}}
\newcommand{\beqs}	{\begin{displaymath}}
\newcommand{\eeqs}	{\end{displaymath}}
\newcommand{\beqas}	{\begin{eqnarray*}}
\newcommand{\eeqas}	{\end{eqnarray*}}

\newcommand{\avg}[1]  {{\langle #1 \rangle}} 
\newcommand{\dis}{\displaystyle}
\newcommand\bit{\begin{itemize}}
\newcommand\eit{\end{itemize}}
\newcommand{\e}	{$^{-1}$}

\newcommand{\eee}	{$^{-3}$}
\newcommand\simlt{\lower.5ex\hbox{$\; \buildrel < \over \sim \;$}}
\newcommand\simgt{\lower.5ex\hbox{$\; \buildrel > \over \sim \;$}}

\font\tenbi=cmmib10 
\newfam\bifam  \textfont\bifam=\tenbi

\font\tenbr=cmbx10
\newfam\brfam  \textfont\brfam=\tenbr

\font\squinttenbi=cmbx10 at 9pt
\scriptfont\brfam=\squinttenbi

\newcommand\vectimes{{\mathbf{\times}}}
\newcommand\vecnabla{
              \setbox1=\hbox{$\bigtriangledown$}
                           \raise.45ex\hbox{$\bigtriangledown$\hskip-.97\wd1
                           $\bigtriangledown$\hskip-.97\wd1
                           $\bigtriangledown$\hskip-.97\wd1}
                           \raise.47ex\hbox{$\bigtriangledown$}}
\newcommand\grad{\vecnabla}
\newcommand\curl{\vecnabla\vectimes}

\newcommand\vecv{{\textbfit{v}}}

\newcommand\vecB{{\textbfit{B}}}

\newcommand\cala{{{\cal A}}}
\newcommand\calm{{{\cal M}}}
\newcommand\caln{{{\cal N}}}
\newcommand\calo{{{\cal O}}}
\newcommand\calr{{{\cal R}}}

\newcommand{\ppbyp}[2]	{{\frac{\partial#1}{\partial#2}}}

\newcommand{\alfven}    	{{Alfv$\acute{\rm e}$n}}

\newcommand\cs		{c_{\rm s}}

\newcommand\ma		{\calm_{\rm A}}

\newcommand{\muh}	{\mu_{\rm H}}

\newcommand{\nh}		{n_{\rm H}}
\newcommand{\nhei}		{n_{{\rm H,\,eq},i}}
\newcommand{\nhm}		{n_{\rm H,\,max}}
\newcommand{\nhn}		{n_{{\rm H},\,\nu}}
\newcommand{\nho}		{n_{{\rm H},0}}
\newcommand\pc		{{\rm pc}}
\newcommand{\rms}       	{{\rm rms}}
\newcommand{\tff}		{t_{\rm ff}}
\newcommand{\tffo}		{t_{\rm ff,0}}
\newcommand{\va}		{v_{\rm A}}
\newcommand{\vao}		{v_{\rm A0}}

\usepackage{color}

\newcommand\damp		{{\rm damp}}
\newcommand\dr		{{\rm dr}}

\newcommand\ead		{\eta_{\rm AD}}
\newcommand{\eb}		{{\cal E}_B}

\newcommand{\ebn}		{{\cal E}_{B_\nu}}
\newcommand{\ebo}		{{\cal E}_{B0}}
\newcommand{\eq}		{{\rm eq}}
\newcommand{\eqi}		{{{\rm eq},i}}
\newcommand\gad		{\gamma_{\rm AD}}
\newcommand\gno		{\Gamma_{\nu 0}}
\newcommand\he		{{\rm He}}
\newcommand\hhp		{{\rm H\,H^+}}
\newcommand\kin		{{\rm kin}}
\newcommand\krho    {{k_\rho}}
\newcommand\lj		{{\lambda_{\rm J}}}
\newcommand\mh		{{m_{\rm H}}}
\newcommand{\non}	{{\rm nl}}
\newcommand\nuni		{\nu_{ni}}

\newcommand\phiff		{\phi_{\rm ff}}
\newcommand\pmn		{P_{m,\,\rm num}}

\newcommand\rmcr		{R_{m,\,\rm cr}}
\newcommand\rtj		{\rho_{\rm TJ}}
\newcommand\sat		{{\rm sat}}

\newcommand\sm		{{\rm sm}}
\newcommand\sph		{{\rm sph}}

\newcommand\vecomega	{{\boldsymbol{\omega}}}
\newcommand\vir		{{\rm vir}}

\newcommand\vtf		{v_{t,5}}
\newcommand\xif		{x_{i,-4}}
\newcommand\ztf		{z_{25}}

\newcommand\asph	{\alpha_{\rm sph}}

\newcommand\kd		{k_{1-\delta}}
\newcommand\kh		{k_{1/2}}
\newcommand\pmg		{P_{m,g}}

\newcommand\tcoll	{{t_{\rm coll}}}

\title[Magnetic Fields in the Formation of First Stars. I]{Magnetic Fields in the Formation of the First Stars. I. Theory vs. Simulation}

\author[C. F. McKee et al.]{
Christopher F. McKee,$^{1,2}$\thanks{E-mail: cmckee@astro.berkeley.edu (CFM)}
Athena Stacy,$^{2}$
and Pak Shing Li$^{2}$
\\
$^{1}$Department of Physics, University of California, Berkeley CA 94720 USA\\
$^{2}$Department of Astronomy, University of California, Berkeley CA 94720 USA\\
}
\date{Accepted XXX. Received YYY; in original form ZZZ}

\pubyear{2020}

\begin{document}
\label{firstpage}
\pagerange{\pageref{firstpage}--\pageref{lastpage}}
\maketitle

\begin{abstract}
While magnetic fields are important in contemporary star formation, their role in primordial star formation is unknown.
Magnetic fields of order $10^{-16}$~G are produced by the Biermann battery due to the curved shocks and turbulence associated with the infall of gas into the dark matter
minihalos that are the sites of formation of the first stars. These fields are rapidly amplified by a small-scale dynamo until they saturate at or near equipartition with the turbulence in the central region of the gas. 
Analytic results are given for the outcome of the dynamo, including the effect of compression in the collapsing gas.
The mass-to-flux ratio in this gas is 2-3 times the critical value, comparable to that in contemporary star formation. Predictions of the outcomes of simulations using smooth particle hydrodynamics (SPH) and grid-based adaptive mesh refinement (AMR) are given. Because the numerical viscosity and resistivity for the standard resolution of 64 cells per Jeans length are several orders of magnitude greater than the physical values, dynamically significant magnetic fields affect a much smaller fraction of the mass in simulations than in reality. An appendix gives an analytic treatment of free-fall collapse, including that in a constant density background. Another appendix presents a new method of estimating the numerical viscosity; results are given for both SPH and grid-based codes.

\end{abstract}
\vspace{0.0in}


\section{Introduction}

The first stars and galaxies were the early drivers of cosmic evolution, directing the universe towards the highly structured state we observe today.  
The radiation emitted during the lifetime of the first stars (a.k.a. primordial, Population III, or Pop III stars), and the metals they released through supernova (SN) explosions and stellar winds, left a crucial imprint on their environment.  
In the wake of Pop III stars, the first galaxies emerged to continue the process of reionizing the universe 
(e.g. \citealt{kitayamaetal2004,syahs2004,whalenetal2004,alvarezetal2006,johnsongreif&bromm2007})
and chemically enriching the intergalactic medium (IGM)(e.g., \citealt{madauferrara&rees2001,
chenetal2017}; reviewed in \citealt{karlssonetal2013}).  
Before Pop III stars first formed, no metals or dust existed to aid in the cooling and condensation of gas into stars.  Primordial star formation was instead driven by cooling through H$_2$ transitions.  Thus, Pop III stars are believed to have initially formed at $z\sim20-30$ in small dark matter halos of mass $\sim10^6$ $M_{\odot}$, since these `minihalos' were the first structures whose constituent gas had a sufficient H$_2$ abundance to allow for star formation (\citealt{haimanetal1996,tegmarketal1997,yahs2003}).

Pop III stars are too faint to be detectable by even next-generation telescopes such as ${\it JWST}$ (\citealt{gardneretal2006}).  Understanding of these objects must instead come from numerical simulations and indirect observational constraints.  
Early studies found that Pop III stars are massive and form in isolation (\citealt{brommetal2002,abeletal2002,bromm&loeb2004,yoh2008}).  More recent work has modified this picture (\citealt{turketal2009, stacyetal2010, stacyetal2012, brom13}):  While the Pop III initial mass function (IMF) is top-heavy, improved simulations have found that a given massive Pop III star forms within a disk and tends to have a number of companions with a range of masses ($\sim$ 1 to several tens of $M_{\odot}$, e.g. \nocite{clarketal2008,turketal2009,clarketal2011a,stacyetal2010,stacyetal2016} \citealp{clarketal2008, clarketal2011a}).

These studies did not include magnetic fields, although magnetic fields have significant effects in contemporary star formation
(see the reviews by \citealp{mckee&ostriker2007} and \citealp{krum19}). Magnetic fields have existed on wide range of astronomical scales for most of the history of the universe (see \citealt{becketal1996,kulsrud&zweibel2008,durreretal2013,subr16} for reviews).  In describing the strength of primordial fields, we sometimes use the
comoving field, $B_c=a^2 B$, where $a=1/(1+z)$ is the cosmological scale factor; this is the value the field would have if it evolved
from redshift $z$ to today under the conditions of flux freezing. 
Primordial magnetic fields could have arisen during inflation, but such fields are extremely small unless the conformal invariance of the electromagnetic field is broken \citep{turn88}. Even in that case, the fields produced are on very small scales and will dissipate unless turbulent motions stretch and fold the field, thereby generating a small-scale dynamo that amplifies the field \citep{durreretal2013}. For example, turbulence driven by primordial density fluctuations drives
a small-scale dynamo acting on inflation-generated seed fields that \citet{wagstaffetal2014} estimate produces fields 
of maximum strength $B_c \sim 10^{-15}$ G 
on comoving scales $\sim 0.1$ pc (under the assumption that they were in equipartition with the turbulence when they were created). 
Magnetic fields can also be produced during an electroweak or QCD phase
transition, although in the standard model these transitions are not first order and do not result in observable fields today \citep{durreretal2013}.
If effects beyond the standard model render one or both these transitions to be first order phase transitions, then they could result in fields of
$10^{-15}-10^{-12}$~G on scales of $0.1-100$~pc today \citep{wagstaffetal2014}. In any case, it is believed that the peak in the field strength
occurs on a scale $L\sim \va/H$, where $L$ is the correlation length of the field, $H$ the Hubble parameter and $\va$ the \alfven\ velocity; the comoving field decreases, and the comoving correlation length increases, with cosmic time, and are now related by $B_c\sim 10^{-14} L_c\,$/(1 pc)~G \citep{banerjee&jedamzik2004}. 
This is only slightly above the observed lower limit on the intergalactic magnetic field of a few times $10^{-15}$~G for correlation lengths of 1 pc based on gamma ray observations of blazars \citep{nero10,tayl11}, although this method of inferring the field has recently been called into question \citep{brod18,alve19}.
A more exotic possibility is that the field results from the chiral magnetic effect
in the epoch of the electroweak transition
due to a difference in the number of left- and right-handed fermions, which \citet{scho18} 
estimate could give a field $B_c\sim 2\times 10^{-16} (L_c/1\;\pc)^{-1/2}$~G. In sum, inflation or phase transitions in the early universe could
generate intergalactic fields as large as $\sim 10^{-13}$~G on scales $\sim 10$~pc
\citep{durreretal2013}; however, these estimates rest on an
uncertain theoretical foundation.

Weaker magnetic fields can definitely be produced through the Biermann battery process, 
in which non-parallel gradients in the electron density and pressure generate
solenoidal electric fields that in turn generate magnetic fields \citep{bier50,bier51}. 
For the Galaxy, these authors estimated that this process would produce a field of order $10^{-19}$~G and that this field
would be subsequently amplified in a turbulent dynamo until it reached approximate equipartition with the turbulent motions. Since the turbulent velocity increases with scale, the magnetic field will also. Research since then has filled in this basic picture \citep{pudr89,kuls97,davi00,xuetal2008}.
Fields created during galaxy formation can be produced in oblique shocks, with an estimated strength $\sim 10^{-18}-10^{-19}$~G \citep{pudr89,xuetal2008}. Weaker fields ($\sim 10^{-24.5}$~G at redshifts $z\sim 10-100$)
can form throughout the universe after recombination due to misalignment of the density gradients in the gas and the temperature gradients in the cosmic background radiation \citep{naoz&narayan2013}.

The small-scale dynamo is also active during the initial collapse of the turbulent gas in cosmic minihalos that leads to the formation of the first stars. Numerical simulations have shown that the field grows due to both a small-scale dynamo and to compression; a resolution of at least 32-64 cells per Jeans length is required to see the operation of the dynamo \citep{suretal2010,federrath11b,turketal2012}. These authors noted that the growth rate of the field increases with the Reynolds number and therefore with resolution; the results were far from converged even at a resolution of 128 cells per Jeans length. A subsequent simulation \citep{koh&wise2016}, which focused on the evolution of the star, its HII region, and the subsequent supernova, found considerably less dynamo amplification. None of these simulations were carried to the point that the field reached approximate equipartition with the turbulent motions prior to the formation of the star. In view of the challenges faced by numerical simulations, semi-analytic approaches have been used to follow the evolution of the field until it saturates: \citet{schleicheretal2010} developed a simple model for the turbulence in a collapsing cloud and the growth of the field, and both they and \citet{schoberetal2012b} used the \citet{kazantsev1968} equation to follow the growth of the field in a turbulent medium. A comprehensive analytic treatment of the small-scale dynamo under conditions appropriate for the formation of the first stars and galaxies has been given by \citet{xu&lazarian2016}.

Magnetic fields can be amplified at later evolutionary times also.  A dynamo driven in a primordial protostellar disk can amplify the field to the point that the magneto-rotational instability (MRI) can operate in the disk, and it can also lead to the generation of outflows and jets (\citealt{tan&blackman2004}). 
Simulations by \cite{machidaetal2006} found that protostellar jets would be launched for initial field strengths of $B > 10^{-9}  \left(n/10^3 {\rm cm}^{-3}  \right)^{2/3}$ G.  
The simulations of \cite{machida&doi2013}, which resolved the gas collapse up to protostellar density and the subsequent evolution for the next few hundred years, found that sufficiently strong magnetic fields 
($>10^{-9}$~G in a Bonnor-Ebert sphere with a central density of $10^4$~cm\eee)
prevented disk formation and led to the formation of a single massive star. However, they did not include the turbulence that has been found to be important in the formation of magnetized disks \citep{gray18}, and their assumption of a uniform initial field is incompatible with having a field of that magnitude being produced by a small-scale dynamo.

\cite{petersetal2014} studied the influence of both magnetic fields and metallicity on primordial gas cut out from cosmologically simuated minihalos, testing metallicities ranging from $Z=0$ to $10^{-4}\,Z_{\odot}$ and initial magnetic fields ranging from zero to $10^{-2}$ G.  They followed their simulations until 3.75 $M_{\odot}$ of gas was converted into star(s), and similarly find multiple sink formation in all cases except for metal-free gas with the largest initial magnetic fields. \citet{shar20} carried out a large number of simulations of primordial star formation with different initial field strengths and found that the magnetic field strongly suppressed fragmentation, thereby significantly reducing the number of low-mass stars that could survive until today. Both groups conclude that
magnetic fields are essential to determining the IMF as well as the binarity and multiplicity of Pop III stars.

This is the first of two papers in which we study the magnitude of the magnetic fields expected in the formation of the first stars and the effects
of these fields on the formation of these stars. As described above, the fields generated either in the early universe or by the
Biermann battery after recombination are very weak, so the fields must be amplified in a small-scale dynamo by a large factor in order
to have an effect on star formation. In this first paper, we review the theory of such dynamos for both the
case in which the dissipation is due to resistivity, which is relevant for numerical simulations, and the case in which
the dissipation is due to ambipolar diffusion, which is relevant for star formation in the epoch between recombination and reionization (Section \ref{sec:ssd}). 
We assume that the initial conditions for the dynamos are set by the Biermann battery operating in the gas that falls into a dark matter minihalo.
We evaluate the quantities that govern the behavior of the dynamos (Table \ref{tab:dyn}) and then include the effects of
gravitational collapse in our analysis. In Section \ref{sec:predthy} we apply these results to the formation of the first stars and
show that magnetic fields can grow to approximate equipartition in the gravitational collapse that forms these stars. It is not currently possible
to carry out simulations with the resolution needed to accurately represent the viscosity and resistivity of the gas that forms
the first stars, so in Section \ref{sec:predsim} we estimate the magnitude of the fields that can be produced by either an
SPH or a grid-based simulation of a small-scale dynamo.
Appendix \ref{app:viscosity} summarizes the values of the viscosity and the ambipolar and Ohmic resistivities under the conditions appropriate for the formation of the first stars. In Appendix \ref{app:free} we describe gravitational collapse in the presence of a fixed dark matter background.
Finally, in Appendix \ref{app:numerical}, we estimate the numerical viscosity for both grid-based and SPH codes, and the resistivity for grid-based codes. In Paper II (Stacy et al in preparation) we simulate the formation of a first star from cosmological initial conditions and compare the results with the theory developed here.

\section{small-scale Dynamos}
\label{sec:ssd}

As noted in the Introduction, the initial cosmological seed field is very weak, but it can be rapidly 
amplified by the small-scale dynamo driven by turbulence \citep{bat50,kazantsev1968,kuls92,sche02a,sche02b,schleicheretal2010,schoberetal2012a,xu&lazarian2016}. Direct experimental evidence for dynamo amplification of magnetic fields in a laser-produced turbulent plasma has been obtained by \citet{tzef18}.
The behavior of the dynamo is set by the
relative sizes of the viscous scale, $\ell_\nu$, where $\nu$ is the kinematic viscosity, and the magnetic dissipation scale, $\ell_\eta$, where $\eta$ is the resistivity \citep{kuls92,schoberetal2012b}. In a fully ionized plasma, $\ell_\eta$ is set by Ohmic resistivity, but in a partially ionized plasma it is generally set by ambipolar diffusion.\footnote{The  ambipolar resistivity as defined by \citet{pint08a} is sometimes termed the magnetic diffusivity.}  The ratio of these scales is determined by
the magnetic Prandtl number, 
\beq
P_m\equiv\frac{\nu}{\eta}.
\label{eq:pm}
\eeq
For Kolmogorov turbulence, $\ell_\nu/\ell_\eta=P_m^{1/2}$ for $P_m\gg 1$ \citep{sche02a} and $\ell_\nu/\ell_\eta=P_m^{3/4}$ for $P_m\ll 1$ \citep{mofa61}.
Most dilute astrophysical plasmas are highly conducting and
have $P_m\gg1$ (e.g., \citealp{sche02a}), so that the resistive scale is small compared to the viscous scale. 
Turbulence both stretches and folds the field. 
The stretching occurs on the eddy scale, and for $P_m\gg 1$ the fastest eddies are on the viscous scale. The eddy motions result in many field reversals, which can survive down to the magnetic dissipation scale.
As a result, the field becomes very anisotropic, varying
on a scale $\ell_\nu\gg\ell_\eta$ parallel to the field and on a scale that decreases in time from $\ell_\nu$ to a scale $\geq \ell_\eta$
normal to the field. In the opposite limit in which $P_m\ll 1$, the field cannot respond to eddies at
the viscous scale, but is instead driven by eddies on the resistive scale. 
In either case, the dynamo is termed ``small-scale," since the field is amplified on scales smaller than
the outer scale of the turbulence.

Since primordial gas cannot cool to very low temperatures, the
turbulence in regions where the first stars form is generally transonic or subsonic, so
for simplicity we shall assume Kolmogorov turbulence in our analytic discussion. The turbulent velocity on a scale
$\ell$ in the inertial range therefore satisfies
$v_\ell\propto \ell^{1/3}$. The quantity $v_\ell^3/\ell$ is then constant in the inertial range and is comparable to the specific energy
dissipation rate, $\epsilon$. Following \citet{pope00}, we define the velocity on the scale $\ell$ as 
\beq
v_\ell\equiv (\epsilon\ell)^{1/3}.
\label{eq:vell}
\eeq
One can show that
then $\frac 12 v_\ell^2\simeq kE(k)$, where $E(k) dk$ is the energy in the range of wavenumbers $dk$. In particular,
$v_\nu=(\epsilon\ell_\nu)^{1/3}$ is the velocity that eddies at the viscous scale, $\ell_\nu$, would have in the absence of dissipation at that scale.
The viscous scale length, $\ell_\nu$, is defined by the condition that the Reynolds number at the scale $\ell_\nu$ is unity,
$Re(\ell_\nu)=\ell_\nu v_\nu/\nu=1$, so that
$\nu=\ell_\nu v_\nu$.
As a result we have 
\beq
\ell_\nu=\left(\frac{\nu^3}{\epsilon}\right)^{1/4},~~~~v_\nu=(\epsilon\nu)^{1/4},
~~~~\Gamma_\nu=\frac{v_\nu}{\ell_\nu}=\left(\frac{\epsilon}{\nu}\right)^{1/2},
\label{eq:ellnu}
\eeq
where $\Gamma_\nu$ is the characteristic eddy turnover rate on the viscous scale. 
The hydrodynamic and magnetic Reynolds numbers of a turbulent flow, $Re$ and $R_m$, depend on the outer scale of the turbulence, $L$:
\beq
Re\equiv\frac{Lv_L}{\nu}=\left(\frac{L}{\ell_\nu}\right)^{4/3}=\left(\frac{v_L}{v_\nu}\right)^4,~~~R_m\equiv\frac{Lv_L}{\eta}=P_m Re.
\label{eq:re}
\eeq


\subsection{Ideal MHD}
\label{sec:ideal}

If the resistivity is negligible, so that $P_m\gg 1$, and if the fluid is incompressible, then in the kinematic limit
the equation for the magnetic energy density
per unit mass, $\eb=B^2/(8\pi\rho)=\frac 12 \va^2$, where $\va$ is the \alfven\ velocity, is \citep{bat50,kuls92}
\beq
\frac{d\eb}{dt}=2\Gamma\eb,
\label{eq:deb}
\eeq
where, as noted above, the growth rate, $\Gamma$, is dominated by eddies on the viscous scale,
\beq
\Gamma=\frac{\avg{\vecB\vecB\bf{:}\vecnabla\vecv}}{\avg{B^2}}\simeq \frac{v_\nu}{\ell_\nu}\equiv\Gamma_\nu,
\label{eq:gamma}
\eeq
and where the angular brackets $\langle\;\rangle$ represent a volume average \citep{sche02b}.
Now, in Kolmogorov turbulence, the eddy turnover rate at the viscous scale is related to that at the outer scale by
\beq
\frac{v_\nu}{\ell_\nu}=\frac{1}{\ell_\nu} \left(\frac{v_L \ell_\nu^{1/3}}{L^{1/3}}\right)=\left(\frac{v_L}{L}\right)Re^{1/2},
\label{eq:vl}
\eeq
where the second step follows from equation (\ref{eq:re}).
\citet{schoberetal2012a} used the WKB approximation to solve
the equation that \citet{kazantsev1968} derived to describe the kinematic dynamo in incompressible,
turbulent fluids and showed that when the resistivity is negligible ($P_m\gg 1$), the growth rate of the field is
\beq
\Gamma=\frac{37}{36}\left(\frac{v_L}{L}\right) Re^{1/2}=\frac{37}{36} \left(\frac{v_\nu}{\ell_\nu}\right)\simeq\Gamma_\nu.
\eeq
In other words, the growth rate is the eddy turnover time at the viscous scale
in this limit. Hence, in the kinematic limit the field energy grows as
\beq
\eb=\ebo e^{2\Gamma_\nu t}.
\label{eq:eb1}
\eeq

On scales larger than the peak of the magnetic power spectrum, the magnetic power spectrum is given by 
\beq
M(k,t)=M_0(k\ell_\nu)^{3/2}e^{\frac 34\int\Gamma_\nu dt}
\eeq
\citep{kazantsev1968,kuls92,sche02b,xu&lazarian2016},\footnote{\citet{kazantsev1968} actually gave a range of exponents for
the wavenumber; \citet{kuls92} appear to have been the first to specify that the exponent is $\frac 32$.}
where we have adopted the normalization of \citet{xu&lazarian2016}.
Under the assumptions that the spectrum varies as $k^{3/2}$ up to the wavenumber at the peak, $k_p$, and then cuts off rapidly \citep{kuls92,xu&lazarian2016}
and that the magnetic energy is initially concentrated at the viscous scale, $k_p\ell_\nu\sim 1$, the energy
in the field is
\beq
\eb(t)=\frac 12\int_0^{k_p}M(k,t)dk=\ebo(k_p\ell_\nu)^{5/2} e^{\frac 34 \int\Gamma_\nu dt},
\label{eq:eb2}
\eeq
where $\ebo=M_0/(5\ell_\nu)$
and we have set $\Gamma=\Gamma_\nu$, as is appropriate for $P_m>1$.
Our normalization for $\eb$ differs by a factor 5 from that adopted by \citet{xu&lazarian2016};
it gives $\ebo=\eb(t=0)$ at $t=0$ for $k_p(t=0)=\ell_\nu^{-1}$.
This relation is valid so long as the dynamo is in the kinematic stage and is driven by eddies at the viscous scale, even in the presence of dissipation,
since the exponential growth occurs on large scales where
dissipation is negligible. In the initial stage of the dynamo, when dissipation is negligible on all relevant scales,
the field energy exponentiates as $\exp(2\Gamma_\nu t)$ (equation \ref{eq:eb1}). It follows from equation (\ref{eq:eb2}) that if the spectrum cuts off sharply for $k>k_p$ in this case, then 
$k_p\propto \exp(\frac 12 \Gamma_\nu t)$. (In fact, the spectrum does not cut off sharply at $k_p$ and the actual peak of the power spectrum evolves as $\exp(\frac 35\Gamma_\nu t)$---\citealp{sche02b}.) As noted above, in the absence of dissipation the field energy is concentrated at a wavenumber $k_p$ that becomes increasingly larger than the viscous scale $\ell_\nu^{-1}$ with time as the eddies wind up the field. 

The subsequent evolution of the field has been discussed by \citet{schoberetal2015}, who considered a range of turbulent Mach numbers such that $v_\ell\propto\ell^\theta$ with $\frac 13\leq\theta\leq\frac 12$, and by \citet{xu&lazarian2016}, who focused on the case of subsonic turbulence ($\theta=\frac 13$) and obtained good agreement with simulations; we shall follow the latter treatment here.
\citet{xu&lazarian2016} pointed out that the exponential amplification slows when the field energy first reaches equipartition with the viscous eddies on the scale $\ell_\nu$, so that
$\eb=\frac 12 v_\nu^2\equiv E_\nu$. The corresponding equipartition field (with $\va^2= v_\nu^2$) is
\beq
B_\nu=(4\pi \rho)^{1/2}v_\nu=(4\pi\rho)^{1/2}(\epsilon\nu)^{1/4}
\label{eq:bnu}
\eeq
from equation (\ref{eq:ellnu}).
In the subsequent transition stage, the turbulent cascade maintains the
viscous-scale eddies while at the same time amplifying the field on successively larger scales until the peak in the magnetic
power spectrum reaches $\ell_\nu^{-1}$. They assume that the energy at the peak (equation \ref{eq:eb2}) remains equal to 
$E_\nu$ during this evolution. The transition stage ends when $k_p\ell_\nu=1$, so that the magnetic forces can stop the
the eddies at that scale.

At this time ($t=t_\non$), the dynamo enters the fully nonlinear stage. Setting $\eb(t_\non)=E_\nu=B_\nu^2/(8\pi\rho)$ for $k_p\ell_\nu=1$ in equation (\ref{eq:eb2}) gives
\beq
t_\non=\frac{4}{3 \Gamma_\nu}\ln\left(\frac{E_\nu}{\ebo}\right)=\frac{8}{3 \Gamma_\nu}\ln\left(\frac{B_\nu}{B_0}\right)~~~~~(P_m\gg 1)
\label{eq:tnl}
\eeq
for the time at which the dynamo enters the fully nonlinear stage.
For example, if the equipartition field at the viscous scale is 10 orders of magnitude above the initial field, then 
this time is $t_\non=61\,\ell_\nu/v_\nu=(3760/Re)^{1/2} L/v_L$. 
Subsequently, it is the smallest eddies that are not suppressed by magnetic forces that dominate the magnetic
energy, so that $\eb\simeq \frac 12 v_\ell^2$ and $\Gamma=\chi v_\ell/\ell$, where $\chi$ is of order unity. It follows that
\beq
\frac{d\eb}{dt}=2\left(\frac{\chi v_\ell}{\ell}\right)\cdot \frac 12 v_\ell^2=\chi\epsilon
\label{eq:deb2}
\eeq
from equation (\ref{eq:deb})  \citep{sche02b}. As a result, the magnetic energy 
in the nonlinear stage is
\beq
\eb=\eb(t_\non)+\chi \epsilon (t-t_\non)~~~~~~~(t>t_\non).
\label{eq:ebnl}
\eeq

\citet{kuls92} presented analytic arguments suggesting $\chi=3/38=0.079$ for the case in which the dissipation
is dominated by reconnection, and \citet{xu&lazarian2016} confirmed this.
Note that in these theories the value of $\chi$ is independent of the rate of reconnection: \cite{kuls92} assumed Petschek reconnection, which has a rate that depends on $R_m$, whereas \citet{xu&lazarian2016} assumed turbulent reconnection, which is maximally efficient and has a rate that is independent of $R_m$.
Numerical simulations confirm that $\chi$ is significantly smaller than unity: \citet{choetal2009} found
$\chi\simeq 0.07$ and \citet{beresnyak2012} found $\chi\simeq 0.05$.
Collectively, these results indicate that 
\beq
\chi^{-1}=16\pm 0.1\mbox{ dex},
\eeq
so we shall adopt $\chi=1/16$ for numerical estimates.
For $t\gg t_\non$, the time to reach equipartition at a scale $\ell$ (i.e., the time at which $\eb=\frac 12 v_\ell^2$) is proportional to
the eddy turnover time,
\beq
t_\eq(\ell)=\frac{\frac 12 v_\ell^2}{\chi\epsilon}=\frac{\ell}{2\chi  v_\ell},
\eeq
so that it takes $(2\chi)^{-1}\sim 8$ eddy turnover times at a scale $\ell$ for the field to reach
equipartition at that scale.

The field stops growing when it reaches equipartition with the largest eddies, $B\simeq B_\eq$, where
\beq
B_\eq=(4\pi\rho)^{1/2}v_L= Re^{1/4}\,B_\nu
\label{eq:beq}
\eeq
from equations (\ref{eq:re}) and  (\ref{eq:bnu}). Simulations suggest that for 
subsonic solenoidal turbulence 
the magnetic field saturates at a value $B_\sat=\phi_\sat B_\eq$ with
$\phi_\sat\simeq (3/7)^{1/2}=0.65$ \citep{haug04} $\simeq 0.7$ \citep{federrath11a,brandenburg14}; for supersonic solenoidal turbulence, \citet{federrath11a}'s results imply $\phi_\sat\simeq 0.14$.

To determine how long it takes for the field to reach equipartition at the scale $L$, we can use
equations (\ref{eq:ellnu}), (\ref{eq:bnu}), and (\ref{eq:tnl}) and the fact that $\eb(t_\non)=B_\nu^2/(8\pi\rho)$ to rewrite equation (\ref{eq:ebnl}) as
\beq
B^2=B_\nu^2\left\{1+2\chi\left[\Gamma_\nu t-\frac{8}{3}\ln\left(\frac{B_\nu}{B_0}\right)\right]\right\}~~~~~~~(t>t_\non).
\eeq
Equation (\ref{eq:beq}) then implies that
\beq
\Gamma_\nu t_{\rm eq}(L)= \frac{8}{3}\ln\left(\frac{B_\nu}{B_0}\right)+\frac{Re^{1/2}-1}{2\chi}\simeq 8 Re^{1/2},
\label{eq:teq}
\eeq
where the final step is for a large Reynolds number and $\chi= 1/16$.
If the field saturates at a value less than $B_\eq$, the factor $Re^{1/2}$ should be multiplied by $\phi_\sat^2$.

There is an aspect of this analysis that is overly idealized: We have assumed that the turbulence is established
instantaneously, whereas in fact it takes at least an eddy turnover time for the turbulence to develop
(e.g., \citealp{banerjee&jedamzik2004}). For a flow that is initialized at some point in time (for example,
at the epoch of recombination), the size of the largest eddy in a turbulent cascade at a time $t$ later is
$L\sim v_L t$. As a result, $\epsilon\sim v_L^2/t$, and equation (\ref{eq:ebnl}) implies
\beq
\frac{\va^2}{v_L^2}\sim 2\chi\left(1-\frac{t_{\non}}{t}\right),
\label{eq:va2}
\eeq
provided $\eb(t_\non)$ is negligible compared to $\eb(t)$. Since $\chi\simeq 1/16$, it follows that the field will be close to equipartition
for $t\gg t_\non$, but can
never reach it unless there is a boundary that sets a limit on $L$, as we implicitly assumed in equation (\ref{eq:teq}).

\subsection{Evolution of the field in the presence of Ohmic resistivity}
\label{sec:resistive}

\begin{figure*}
\includegraphics[width=0.55\textwidth]{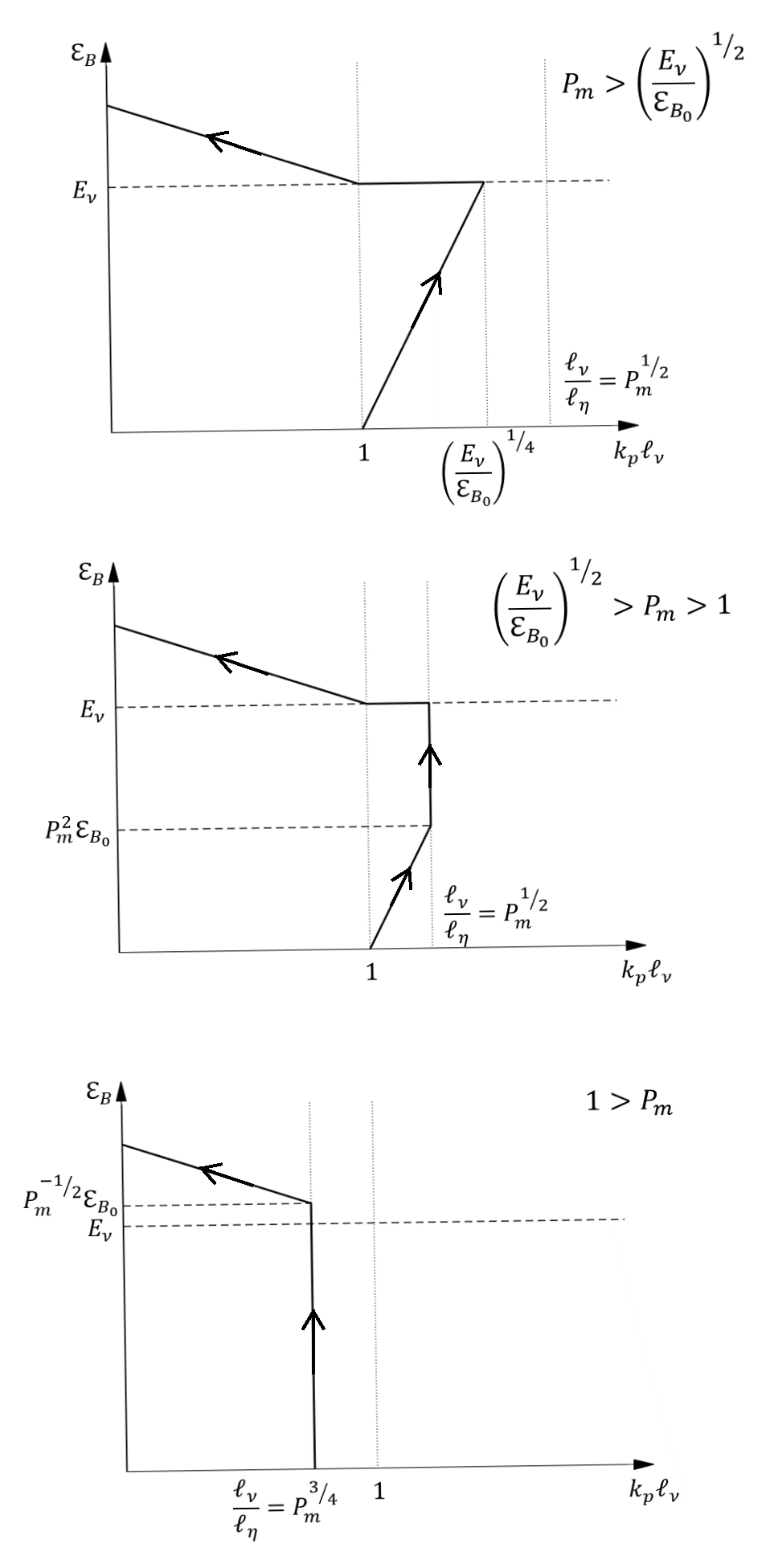}
\caption{
\label{fig:ohm}
Graphical representation of the theory of \citet{xu&lazarian2016} for dynamos with Ohmic resistivity. 
The magnetic specific energy, $\eb$, which increases as the dynamo operates, is plotted against the wavenumber at which the magnetic
power spectrum peaks, $k_p$, normalized by the viscous length scale, $\ell_\nu$. Arrows indicate the direction of time. For very large values of the magnetic Prandtl number,
$P_m$ (top panel), $k_p$ increases until the magnetic energy reaches equipartition with the viscous-scale eddies ($\eb=E_\nu$); $\eb$ then remains 
approximately constant as the energy moves to larger scales. Once the peak wavenumber reaches the viscous scale, the energy resumes
its growth as it taps the energy of larger eddies. For intermediate $P_m$ (middle panel), the increase in $k_p$ in the kinematic stage stops when
$k_p$ reaches the resistive scale. For $P_m<1$ (bottom panel), the peak wavenumber is capped at $k_p\ell_\eta\sim 1$, and nonlinear growth does not
begin until the field reaches equipartition with the turbulence at that scale.
}
\end{figure*}

The evolution of the field in the presence of Ohmic resistivity, in both the kinematic and nonlinear phases, has been worked out
by \citet{xu&lazarian2016}, and we summarize their results in Figure \ref{fig:ohm}. The magnetic specific energy, $\eb$, increases monotonically
with time, whereas the wavenumber at the peak of the magnetic power spectrum, $k_p$, initially increases with time for $P_m>1$; in the nonlinear
stage, $k_p$ decreases with time for all $P_m$. Resistivity has no effect on the dynamo if it is sufficiently small,
it affects
the later part of the kinematic stage of the dynamo for intermediate values of $P_m$, and it delays the onset of the nonlinear stage of the 
dynamo for $P_m<1$. The change in the evolution that is apparent in Fig. {\ref{fig:ohm} as one moves from top to bottom is due to the 
resistive scale, $\ell_\eta$, which is represented by the rightmost vertical line, moving from right to left as $P_m$ decreases.
The resistive scale is too small to matter in the top panel, 
and the dynamo evolves as described above for ideal MHD. 
For intermediate values of $P_m$ (the middle panel), resistivity prevents the peak wavenumber from growing past the inverse of the resistive scale, $\ell_\eta^{-1}$. When the peak wavenumber is fixed due to resistive dissipation, the growth of the specific magnetic energy becomes 
\beq
\eb=\ebo P_m^{5/4}  e^{\frac 34 \int \Gamma_\nu dt}~~~~~[1<P_m<(E_\nu/\ebo)^{1/2}]
\eeq
(\citealp{xu&lazarian2016}; see equation \ref{eq:eb2}).
Finally, for $P_m<1$ (the bottom panel),
the peak in the energy spectrum remains at $\ell_\eta^{-1}$ in the kinematic stage. 
Since $\ell_\eta>\ell_\nu$, the damping scale is
in the turbulent cascade, and the eddy turnover rate at the dissipation scale is given by
equation (\ref{eq:ellnu}) with $\nu$ replaced by $\eta$ (e.g., \citealt{xu&lazarian2016}),
\beq
\Gamma_\eta=\left(\frac{\epsilon}{\eta}\right)^{1/2}=P_m^{1/2}\Gamma_\nu.
\label{eq:geta}
\eeq
The value of the field energy is given by
equation (\ref{eq:eb2}) with $\nu$ replaced by $\eta$ and $k_p\ell_\eta=1$,
\beq
\eb=\ebo e^{\frac 34 \int \Gamma_\eta dt}~~~~~(P_m<1).
\label{eq:ebeta}
\eeq

The condition for the dynamo to enter the nonlinear stage is that the field energy equal the kinetic energy of the eddies driving the dynamo. For $P_m>1$, these eddies are at the viscous scale, and the dynamo enters the nonlinear stage at the time given in equation (\ref{eq:tnl}). For $P_m<1$, so that $\ell_\eta>\ell_\nu$, these eddies are at the resistive scale,
and the dynamo enters the nonlinear stage
at the time given by equation (\ref{eq:tnl}) with $\Gamma_\nu$ replaced by $\Gamma_\eta$ 
and $B_\nu$ replaced by $B_\eta=P_m^{1/4}B_\nu$ \citep{xu&lazarian2016}. 

In Paper II, we address the evolution of the magnetic field with an SPH code ({\sc gadget-2}) that can follow the evolution 
of the kinematic dynamo and a grid-based code ({\sc orion2}) that has full ideal MHD. Neither treats ambipolar diffusion; 
both have numerical resistivity. 
\citet{les07} have argued that grid-based codes have a numerical magnetic Prandtl number,
$\pmn$, between 1 and 2, depending on wavenumber.
In Appendix \ref{app:numerical}, 
we analyze the results of \citet{federrath11b} and conclude that $P_m\simeq 1.4$ for grid-based codes, in good agreement with the result of \citet{les07}.
We adopt the same value of $P_m$ for SPH codes.

In order for the dynamo to operate, it is necessary for the magnetic Reynolds number to exceed a critical value,
$\rmcr$. 
Using numerical simulations, \citet{haug04} found 
\beq
\rmcr\simeq 2\pi\times35 P_m^{-1/2}=220 P_m^{-1/2}~~~~~(0.1\la P_m\la 3),
\label{eq:rmcr}
\eeq
where the factor $2\pi$ has been inserted in order to convert the expression for the Reynolds number used by \citet{haug04}, $R_m=v/(k_f\eta)$, where $k_f=2\pi/L$ is the forcing wavenumber, to the expression adopted here,
$R_m=vL/\eta$.
\citet{haug04} found that $\rmcr$ begins to increase with $P_m$ somewhere beyond
$P_m=3$, reaching 220 at $P_m=10$. 
\citet{schoberetal2012a} solved the
Kazantsev equation in the WKB approximation and found $\rmcr\simeq 107$ for $P_m\gg 1$.
For supersonic turbulence, \citet{fede14} found $\rmcr\simeq 129$, based on large part on simulations with $P_m=10$.
Since simulations of the formation of the first stars are characterized by transonic turbulence and modest values of $P_m$, the results of \citet{haug04} are most relevant for our problem, and we shall adopt the value of $\rmcr$ in equation (\ref{eq:rmcr}) here.

\subsection{Evolution of the field in the presence of ambipolar diffusion}
\label{sec:ambi}

The first stars form in a 
weakly ionized plasma in which the dominant resistivity is
ambipolar diffusion \citep{kuls92,schoberetal2012b,xu&lazarian2016}. For the case of weak ionization ($\rho_i\ll \rho_n\simeq\rho$),
where $\rho_n$ and $\rho_i$ are the neutral and ion mass densities, the resistivity due to ambipolar diffusion is
(e.g., \citealp{pint08a})
\beq
\ead=\frac{B^2}{4\pi\gad\rho_i\rho_n}=\frac{\va^2}{\gad\rho_i}=\frac{\va^2}{\nuni},
\label{eq:ead}
\eeq
where $\gad$ is the collisional drag coefficient and $\nuni$ is the neutral-ion collision frequency (see Appendix \ref{app:viscosity}). 
It follows that $\ead\propto \va^2\propto B^2$, so that the magnetic Prandtl number, $P_m=\nu/\eta$, starts off very large
when evaluated for the primordial field, but then decreases exponentially in time as the small-scale dynamo
amplifies the field. The damping rate of magnetic fluctuations due to ambipolar diffusion is \citep{kuls92}
\beq
\omega_d=\frac 13\,\frac{k^2\eb}{\nuni},
\label{eq:omegad}
\eeq
where the factor $\frac 13$ comes from averaging the rate over angle.

The growth of the magnetic field in the presence of ambipolar diffusion has been analyzed by \citet{kuls92} and, in more detail,
by \citet{xu&lazarian2016}; we follow the latter treatment here
 (see Fig. \ref{fig:AD}, which summarizes their results). The first, dissipation-free stage of the kinematic dynamo has been described in \S \ref{sec:ideal} above.  
Damping is important at the wavenumber, $k_d$, 
at which the damping rate 
equals the rate at which the field is being stretched, $\omega_d(k_d)=\Gamma_\nu$, where it has been assumed
that the field is weak enough that $k_d\ell_\nu>1$ so that the driving is at the viscous scale. As a result, equation (\ref{eq:omegad}) implies
\beq
k_d\ell_\nu=\left(\frac{3\nuni\ell_\nu^2\Gamma_\nu}{\eb}\right)^{1/2}=\left(\frac{\calr E_\nu}{\eb}\right)^{1/2},
\label{eq:kd}
\eeq
where the parameter 
\beq
\calr\equiv \frac{6\nuni}{\Gamma_\nu}
\label{eq:calr}
\eeq
plays a role for the case of ambipolar diffusion similar to that $P_m$ plays
in the resistive case. Since $\calr\propto\nuni$, it varies linearly with the degree of ionization; we therefore term it the ``dynamo ionization parameter." 
We can relate it to the magnetic Prandtl number as follows:
Since $\ead\propto \va^2\propto B^2$, we have $P_m\propto B^{-2}$.
For $B=B_\nu$ -- i.e., when the field energy is in equipartition with the viscous-scale eddies -- we have $\va^2=v_\nu^2$ so that
\beq
P_m(B_\nu)=\frac{\nu}{\ead(B_\nu)}=\frac{\nu}{v_\nu^2/\nuni}=\frac{\nuni}{\Gamma_\nu}=\frac 16 \calr.
\eeq
Hence $\calr$ is a measure of the magnetic Prandtl number when $B=B_\nu$.
If $\calr$ is not too large ($\calr\la (E_\nu/\ebo)^{1/2}$), the kinematic dynamo enters a dissipative stage of evolution in which the peak of the magnetic
energy spectrum is at the damping wavenumber, $k_p=k_d$,
and one finds from equations (\ref{eq:eb2}) and (\ref{eq:kd}) that the magnetic energy grows as $\eb\propto \exp(\Gamma_\nu t/3)$. 
If $\calr>1$ (middle panel of Fig. \ref{fig:AD}), equation (\ref{eq:kd}) shows that
$k_p=k_d$ exceeds $\ell_\nu^{-1}$ when equipartition 
is reached at $\ell_\nu^{-1}$ (i.e., when $\eb=E_\nu$). As in the ideal case, the system then undergoes a transitional stage in which $k_p$ drops to $\ell_\nu^{-1}$
while $\eb=E_\nu$. The transitional stage ends and the nonlinear stage begins
at $t_\non$ given by equation (\ref{eq:tnl}). On the other hand,
for $\calr<1$ (bottom panel of Fig. \ref{fig:AD}), the first dissipative stage ends when $k_d$ drops to
$\ell_\nu^{-1}$, which occurs prior to equipartition according to equation (\ref{eq:kd}). \citet{xu&lazarian2016} showed 
and \citet{xu19} confirmed computationally that subsequently
the magnetic energy grows as $\eb\propto t^2$ 
for a time interval 
\beq
\Delta t_\damp= \frac{23}{3\Gamma_\nu}\left(\frac{1}{\calr}-1\right), 
\label{eq:delta}
\eeq
so that the dynamo enters
the fully nonlinear stage at a time $ t_\non+\Delta t_\damp$, where $t_\non$ 
is given in equation (\ref{eq:tnl}). As in the case of Ohmic resistivity, transition from the case of very high $\calr$ in the top panel of Fig. \ref{fig:AD} 
to low $\calr$ in the bottom panel can 
be visualized as the effects of the line representing $\eb(k_d)$, no longer vertical, sweeping from right to left as $\calr$ decreases.

\begin{figure*}
\includegraphics[width=0.98\textwidth]{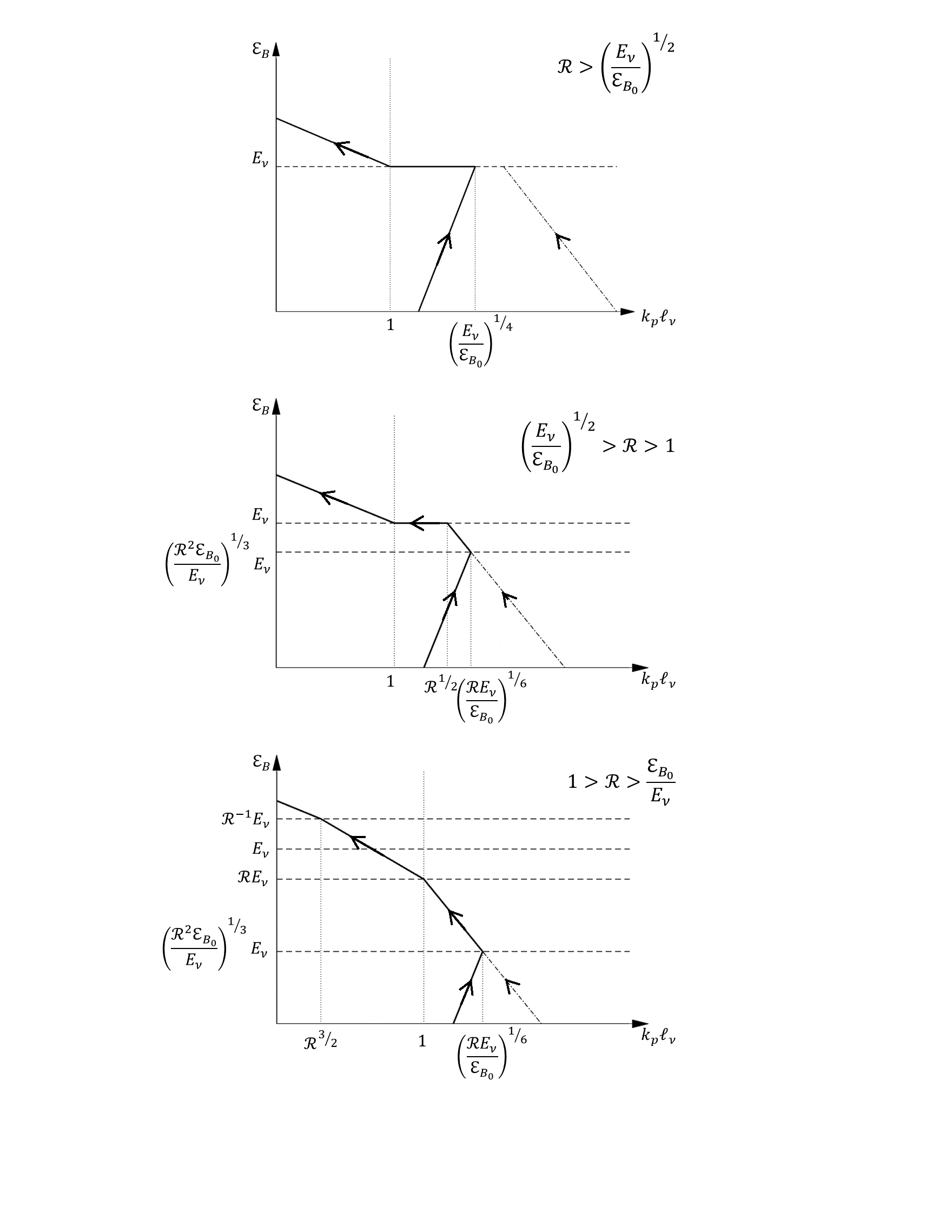}
\vspace{-2.3cm}
\caption{
\label{fig:AD}
Graphical representation of the theory of \citet{xu&lazarian2016} for dynamos in the presence of ambipolar diffusion. 
The magnetic specific energy, $\eb$, which increases as the dynamo operates, is plotted against the wavenumber at which the magnetic
power spectrum peaks, $k_p$, normalized by the viscous length scale, $\ell_\nu$.
The zero is suppressed: $\eb$ begins at $\ebo$ for $k_p\ell_\nu=1$. The damping wavenumber, $k_d$ (equation \ref{eq:kd}, dot-dash line), decreases
as the magnetic energy increases. Arrows indicate the direction of time. For large values of the dynamo ionization parameter,
$\calr$ (equation \ref{eq:calr}; top panel), $k_p$ increases until the magnetic energy reaches equipartition with the viscous-scale eddies ($\eb=E_\nu$). 
For intermediate $\calr$ (middle panel), the damping scale $k_d^{-1}$ becomes large
enough that it determines $k_p$ in the later parts of the kinematic stage. 
For $\calr<1$ (bottom panel), the damping is strong enough that the magnetic specific energy is less than that of the viscous eddies ($\eb=\calr E_\nu < E_\nu$) when
the damping scale grows to the viscous scale. Thereafter, $\eb$ grows as $t^2$ until the field reaches equipartition with the eddies at $k_p$,
when $\eb=\calr^{-1}E_\nu$. In each case, the leftmost stage of evolution is the same as that for Ohmic resistivity.
}
\end{figure*}

To gain more insight into the different stages of the dynamo, one can evaluate
 the magnetic Reynolds number at the dynamo driving scale, $\ell_\dr$.
 With the aid of equation (\ref{eq:ead}) we obtain
\beq
R_m(\ell_\dr)=\frac{\calr}{6}\left(\frac{\Gamma_\nu}{v_\dr/\ell_\dr}\right)\frac{v_\dr^2}{\va^2}.
\eeq
If the driving is at the viscous scale ($\ell_\dr=\ell_\nu$), we have $v_\dr/\ell_\dr=\Gamma_\nu$ so that $R_m(\ell_\dr)>\frac 16 \calr$ in the kinematic stage
($\va^2<v_\dr^2$). For $\calr\geq1$, the dynamo enters the nonlinear stage at $R_m(\ell_\nu)=\frac 16 \calr$. For $\calr<1$, one can use
the results of \citet{xu&lazarian2016} to show that $R_m=\frac 16$ in the damping stage.

We summarize the parameters describing the growth of the magnetic field when ambipolar diffusion dominates in Table \ref{tab:dyn}.
The values of the viscosity, $\nu$, and the ambipolar resistivity, $\ead$, are given in Appendix \ref{app:viscosity}.  
Before applying the results in this table, we first consider the origin of the field and the effect of a time-dependent background on the dynamo.

\begin{table*}
\caption{Turbulent, Ambipolar-Diffusion Dominated Dynamo in Weakly Ionized Plasma in a Cosmic Minihalo
\label{tab:dyn}
}
\begin{tabular}{lcl}
\hline
~~Parameter~~&
~~Equation~~&
~~~~~~~Evaluation$^{a}$~~~~~\\
\hline
$\epsilon=\dis\frac{v_\ell^3}{\ell}\rightarrow \dis\frac{v^3(r)}{r}$&--&~~~~~~$3.20\times 10^{-6}\,\left(\dis\frac{\vtf^3}{r_2}\right)$~~~cm$^2$~s\eee
\vspace{0.3 cm}\\
$\ell_\nu=\left(\dis\frac{\nu^3}{\epsilon}\right)^{1/4}$ &(\ref{eq:ellnu})&~~~~~ $1.42\times 10^{16}\left(\dis\frac{T_3^{0.63}r_2^{1/4}}{\vtf^{3/4}\nh^{3/4}}
\right)$
~~~cm\vspace{0.3 cm}\\
$v_\nu=\left(\epsilon\nu\right)^{1/4}$&(\ref{eq:ellnu})&~~~~~~$3.59\times 10^3\left(\dis\frac{T_3^{0.21}\vtf^{3/4}}{r_2^{1/4}\nh^{1/4}}\right)$~~~cm s\e \vspace{0.3 cm}\\
$\Gamma_\nu=\dis\frac{v_\nu}{\ell_\nu}=\left(\dis\frac{\epsilon}{\nu}\right)^{1/2}$&(\ref{eq:ellnu})&~~~~~$2.52\times 10^{-13}
\,\left(\dis\frac{\vtf^{3/2}\nh^{1/2}}{ T_3^{0.42}r_2^{1/2}}\right)$~~~s\e\vspace{0.3 cm}\\
$B_\nu=(4\pi\rho_\nu)^{1/2}v_\nu$&(\ref{eq:bnu})&~~~~~$1.90\times 10^{-8}\left(\dis\frac{\vtf^{3/4}T_3^{0.21}\nhn^{1/4}}{r_2^{1/4}}\right)$~~~G 
\vspace{0.3 cm}\\
$B_\eq=(4\pi\rho_\eq)^{1/2}v_t$& (\ref{eq:beq})  &~~~~~$5.30\times 10^{-7}\,\vtf\nhei^{1/2}$~~~G\vspace{0.3 cm}\\
$t_\non=\dis\frac{8}{3\Gamma_\nu}\ln\left[\left(\frac{\rho_0}{\rho_\nu}\right)^{\frac 23}\frac{B_\nu}{B_0}\right]^{\mbox~~ b}$&
(\ref{eq:tnl})&~~~~~$\dis 3.35\times 10^5\left(\frac{T_3^{0.42}r_2^{1/2}}{\vtf^{3/2}\nhn^{1/2}}\right)\ln\left[\left(\frac{n_{\rm H,0}}{\nhn}\right)^{\frac 23}\frac{B_\nu}{B_0}\right]$~~~yr
\vspace{0.3 cm}\\
$P_m=\dis\frac{\nu}{\ead}=\frac{R_m}{Re}$&(\ref{eq:pm})
&~~~~~$1.29\left(\dis\frac{\phi_d\xif T_3^{0.80}r_2^{1/2}\nh^{1/2}}{\vtf^{3/2}}\right)\left(\dis\frac{B_\nu^2}{B^2}\right)$
\vspace{0.3 cm}  \\
$\calr=\dis\frac{6\nuni}{\Gamma_\nu}=6\left(\dis\frac{B^2}{B_\nu^2}\right) P_m$&(\ref{eq:kd})&~~~~~$7.72\left(\dis\frac{\phi_d\xif T_3^{0.80}r_2^{1/2}\nh^{1/2}}{\vtf^{3/2}}\right)$
\vspace{0.3 cm}\\
$Re=\dis\frac{Lv_L}{\nu}\rightarrow\dis\frac{rv(r)}{\nu}$&(\ref{eq:re})&~~~~~$6.04\times 10^5\left(\dis\frac{\vtf r_2\nh}{T_3^{0.84}}\right)$
\vspace{0.3 cm}\\
$R_m=\dis\frac{Lv_L}{\ead}\rightarrow\dis\frac{rv(r)}{\ead}$&(\ref{eq:re})&~~~~~$1.00\times 10^3\left(\dis\frac{\phi_d\xif T_3^{0.38}r_2\nh}{\vtf}\right)
\left(\dis\frac{B_\eq^2}{B^2}\right)$\\
\hline
\end{tabular}
\begin{tablenotes}
\item$^{a}${$r_2$ is the outer scale of the turbulence in units of $10^2$ pc, $\vtf$ is the turbulent velocity on that scale in units
of $10^5$ cm s\e, $\nh$ is the density of hydrogen in cm\eee,} $T_3=T/(10^3$~K), $\xif=x_i/10^{-4}$ is the normalized ionization fraction,
and $\rho_\nu$ and $\rho_\eq$ are the densities at $B=B_\nu$ (equation \ref{eq:bnu1.5}), $B_\eq$ (equation \ref{eq:beqi}), respectively.
$\phi_d$ (equation \ref{eq:phid}) measures the importance of the ion-neutral drift velocity; $\phi_d=1$ for $v_d=0$ and $\phi_d\propto (v_d/\cs)^{3/4}$ for highly supersonic drift. We assume $n_{\rm He}/\nh=1/12$ so that $\muh\equiv\rho/\nh=1.33\mh$.
\item$^{b}${Assumes no dissipation and that $P_m\ga 1$ for Ohmic resistivity and $\calr\ga 1$ if the resistivity is due to ambipolar diffusion.}
\end{tablenotes}
\end{table*}

\subsection{The Biermann Battery in a Turbulent Medium}
\label{sec:bier}

As shown by \citet{bier50} (see also \citealp{bier51}), magnetic fields can be generated in an accelerating plasma, a mechanism referred to as the ``Biermann battery." 
An electric field arises in such a plasma in order to maintain charge neutrality if the force per unit mass on the electrons differs from that on the ions.
If the velocity field has a curl, so will the electric field, which produces a magnetic field by Faraday's law.
These authors estimated the magnetic field by noting that the electric field is of order
$E\sim (\mh/e) dv/dt\sim (\mh/e) v^2/\ell$, so that $\partial B/\partial t\sim cE/\ell \sim (c\mh/e) v^2/\ell^2$ and $B\sim (c\mh/e)v/\ell=1.0\times 10^{-4}(v/\ell)$. As noted in the Introduction, they estimated that this process would produce a field of order $10^{-19}$~G in a galaxy.

\citet{harr69,harr70} gave a more rigorous derivation of this result for the case in which the force is radiation drag on the electrons, and
\citet{kuls97} did so for the case in which the force is due to a pressure gradient. The latter authors pointed out
that the equation for the vorticity and that for the magnetic field have the same form,
\beqa
\ppbyp{\vecomega}{t}-\curl(\vecv\vectimes\vecomega)\hspace{-0.3cm}&=&\hspace{-0.3cm}\frac{\grad\rho\vectimes\grad p}{\rho^2}+\nu\nabla^2\vecomega,
\label{eq:dom}\\
\ppbyp{\vecB}{t}-\grad\vectimes\left(\vecv\vectimes\vecB\right)\hspace{-0.3cm}&=&\hspace{-0.3cm}-\frac{m_a c}{e(1+\chi)}\left(\frac{\grad \rho\vectimes\grad p}
{\rho^2}\right)+\eta\nabla^2\vecB,~~~
\label{eq:dbbc}
\eeqa
where $m_a=\rho/n_a$ is the mean mass of the atoms (both neutral and ionized), $n_a$ is the number density
of atoms, and $\chi\equiv n_e/n_a$ is the ionization fraction. These equations are based on the assumption that
$\chi$, $\nu$ and $\eta$ are constant. The source for $\vecomega$ and $\vecB$ is the baroclinic term due to non-parallel density and pressure gradients
($\grad\rho\vectimes\grad p\neq 0$), which arise naturally in curved shocks.

\citet{kuls97} stated that the viscous and resistive terms in equations (\ref{eq:dom}) and (\ref{eq:dbbc}) can be ignored in determining the postshock vorticity. To see this for the viscous term, for example, go into the shock frame, so that $\partial/\partial t =0$, and integrate equation (\ref{eq:dom}) across the shock front. Writing
$p=\rho\cs^2$, where $\cs$ is the isothermal sound speed, we find
\beq
\Delta (v\omega) \sim \frac{\Delta (\cs^2\ln\rho)}{L} +\nu\Delta(\grad\omega),
\eeq
where we have assumed that the vectors in equation (\ref{eq:dom}) are not nearly parallel and where $L$ is the scale of the curvature of the shock. The post-shock sound speed is of order the shock velocity, $v_s$, so the first term on the RHS is of order $v_s^2/L$. The vorticity generated by the shock is of order $v_s/L$. The turbulent cascade behind the shock begins on the scale $L$, so the vorticity changes on that scale just behind the shock; as a result the second term is of order $\nu v_s/L^2$. It follows that the ratio of the first term to the second is of order $v_s L/\nu=Re\gg 1$, so the viscous term does not affect the generation of vorticity in the shock. A similar argument can be made for the evolution of the magnetic field provided that the shock is collisional, as it should be at low velocities in a primarily neutral medium.

It follows that if the vorticity and field are initially zero, they will grow in tandem; for the case in which the force is a pressure gradient, the field is
\beq
\vecB=-\left[\frac{m_a c}{(1+\chi)e}\right]\vecomega.
\label{eq:bom}
\eeq
If the force is due to radiation drag on the electrons, the field is
$\vecB=-(m_a c/e)\vecomega$ in a fully ionized plasma \citep{harr69};
if the plasma is partially ionized, one can show that
the field is larger by a factor $\chi^{-1}$.
\citet{bal93} showed that fields generated by the Biermann battery are 
so weak that the Larmor radius,
$r_{{\rm L},i}=v_{\rm ion}/\Omega_a\simeq(v_{\rm ion}/v_\ell)\ell$,
can exceed the scale $\ell$ on which the vorticity is measured; 
here $v_{\rm ion}$ is the velocity of an individual ion, whereas $v_\ell$ is the mean velocity on the scale $\ell$ and is less
than $v_{\rm ion}$ for subsonic flows.

Numerically, for a vorticity $\omega=v_t(r)/r$ and for $n_{\rm He}=\nh/12$, this field is
\beq
B=1.29\times 10^{-4}\omega=4.17\times 10^{-19}\left(\frac{v_{t,5}}{r_2}\right)~~~\mbox{G},
\eeq
where $\vtf$ is the turbulent velocity in units of $10^5$~cm~s\e\ and $r_2$ is the radius in units of 100 pc.
Although very weak fields ($\sim 10^{-24.5}$~G) can be generated within linear perturbations in the post-recombination universe \citep{naoz&narayan2013}, significantly stronger fields are generated in curved shocks associated with galaxy formation \citep{pudr89} and the accretion of gas into minihalos.

Turbulence leads to an increase in the field in two separate stages, the turbulent Biermann battery and then the small-scale dynamo. First, since the post-shock flow is at high $Re$ (Table \ref{tab:dyn}),
the vorticity on a scale $L$ leads to a turbulent cascade
in which the vorticity increases in time as it cascades to smaller and smaller scales, $\omega\sim v_\ell/\ell\simeq (L/\ell)^{2/3}v_L/L$. 
Correspondingly, the magnetic field increases on smaller scales according
to equation (\ref{eq:bom}) \citep{kuls05}. For $P_m>1$, this process ceases when viscous damping terminates the turbulent cascade on the scale
$\ell_\nu$. The vorticity on this scale is $\sim\Gamma_\nu$, so that the field due to a turbulent Biermann battery is
\beq
B=3.24\times 10^{-17}\left(\frac{\vtf^{3/2}\nh^{1/2}}{T_3^{0.42}r_2^{1/2}}\right)~~~\mbox{G}
\label{eq:turbbier}
\eeq
at the end of this process (see Table \ref{tab:dyn}). 

Once the turbulent cascade has been established, in a time of order $L/v_L$, the vorticity no longer grows and the growth of the field is due to a small-scale dynamo as discussed above.
Here the difference between equations (\ref{eq:dom}) and (\ref{eq:dbbc}) becomes important: $\vecomega=\curl\vecv$ is a function of $\vecv$,
whereas $\vecB$ is not. Thus, while the vorticity no longer grows once the turbulent cascade is established, the magnetic field can
grow exponentially. 

\subsection{Dynamos in a Time Dependent Background}
\label{sec:time}

To this point, we have assumed that the dynamo is operating in a medium with a density that is independent of time.
However, the gas that forms a primordial star first expands with the cosmological expansion, contracts with the formation 
of a minihalo, and then contracts further as it forms a protostellar core. As a result,
the evolution equations for the small-scale dynamo must be revised to account for the temporal evolution of the mean density.
For homologous expansion or collapse, mass and flux conservation imply that $\rho\propto 1/r^3$ and $B\propto 1/r^2$, 
where $r$ is the distance from an arbitrary point in a homologous expansion or from the center of the collapse, which is assumed to be spherical.
As a result, $B\propto \rho^{2/3}$. Collapse is generally
not homologous, so these relations need not hold locally. Nonetheless, prior to the formation of a star, the mean
density and mean field satisfy $\bar B\propto \bar\rho^{2/3}$ under the conditions of flux-freezing. 
\citet{laza15} and references therein argue that reconnection in a turbulent medium leads to violations of flux-freezing,
and \citet{li15} found evidence for this in their simulations. Those same simulations found that this was a modest effect, however,
and were consistent with an overall dependence
$\eb\propto B^2/\rho\propto (\rho/\rho_0)^{1/3}=\xi^{1/3}$.
Following \citet{schleicheretal2010} and \citet{schoberetal2012b}, we assume that the effects of the dynamo and the time dependent background are separable.
As a result, equations (\ref{eq:eb2}) and (\ref{eq:ebeta}) for the kinematic dynamo become 
\beqa
\eb&=&\ebo\xi^{1/3} (k_p\ell_\nu)^{5/2}e^{\frac 34 \int\Gamma_\nu dt}~~~~(P_m>1),
\label{eq:eb3}\vspace{0.2cm}\\
&=& \ebo\xi^{1/3}e^{\frac 34 \int\Gamma_\eta dt}~~~~~~~~~~~~~~~~(P_m<1).
\label{eq:eb3eta}
\eeqa
where 
\beq
\xi\equiv\frac{\rho}{\rho_0}
\end{equation}
is the compression ratio and $\rho_0$ is the initial density.
After a star forms, these equations need not hold, since the mean gas density no longer varies as $1/r^3$
and the magnetic flux released from the star can evolve in a complex manner.

Recall that the dynamo enters the nonlinear stage when $\eb=E_\nu$, the specific energy of the viscous-scale eddies, and also that
$k_p\ell_\nu=1$ at this time. (If ambipolar diffusion dominates, the case in which $\calr<1$ is more complicated as discussed in Section
\ref{sec:ambi}, so we do not discuss that case in this section.)  Let $\rho_\nu$ be the density at the time that the dynamo enters the
nonlinear stage, and let $\avg{\Gamma_\nu}$ be the time-averaged value of $\Gamma_\nu$ prior to that time.
Expressing $\eb$ in terms of $B$, we then find that
the dynamo enters the nonlinear stage at
\beq
t_\non\simeq \frac{8}{3\avg{\Gamma_\nu}}\ln\left(\xi_\nu^{-2/3}\,\frac{B_\nu}{B_0}\right),
\label{eq:tnl2}
\eeq
where $\xi_\nu\equiv\rho_\nu/\rho_0$.
As we shall see in Section \ref{sec:predthy}, $t_\non$ is expected to be small compared to the dynamical time in the formation
of the first stars, so the factor
$\xi_\nu$ in equation (\ref{eq:tnl2}) is close to unity and
$\avg{\Gamma_\nu}\simeq \gno$, the initial value of $\Gamma_\nu$. However, this is not the case for the simulations (Section \ref{sec:predsim}),

For the nonlinear dynamo ($t>t_\non$), equation (\ref{eq:deb2}) becomes
\beq
\frac{d\eb}{dt}=\chi\epsilon+\eb\,\frac{d\ln\xi^{1/3}}{dt},
\label{eq:deb4}
\eeq
where we have assumed that the field has not reached equipartition with motions
on the outer scale of the turbulence ($B<B_\eq$).
The scale
of the dynamo enters through $\epsilon=v_t^3/\ell$.  Equation (\ref{eq:deb4}) then gives
\beq
\eb(t)=\left(\frac{\xi}{\xi_\nu}\right)^{1/3}\ebn +  \chi \xi^{1/3} \int_{t_\non}^t 
\epsilon(t')\xi(t')^{-1/3} dt',
\label{eq:eb5}
\eeq
where $\ebn=\eb(t_\non)$ is given by
\beq
\ebn\equiv \frac{B_\nu^2}{8\pi\rho_\nu}=\frac 12 v_\nu^2=\frac 12 (\epsilon\nu)^{1/2}=\frac 12\,\frac{\epsilon}{\Gamma_\nu}.
\label{eq:ebn1}
\eeq
The first term in equation (\ref{eq:eb5}) represents the compression (assuming the density is increasing) of the field at the beginning of the nonlinear stage ($B_\nu$), whereas the second term represents the field produced by the nonlinear dynamo, including the amplification of that field due to compression.

We approximate the density dependence of a quantity $x$ as $x\propto \rho^{q_x}\propto \xi^{q_x}$. In particular,
$\epsilon\propto\rho^{q_\epsilon}$ and $\Gamma\propto\rho^{q_\Gamma}$, so that
\beq
\eb(t)=\frac 12\left(\frac{\xi}{\xi_\nu}\right)^{\frac 13}\frac{\epsilon_0}{\Gamma_{\nu 0}}\,\xi_\nu^{q_\epsilon-q_\Gamma}+\phiff\chi\epsilon_0\tffo\xi^{\frac 13}I_{q_\epsilon-\frac 13}(\xi_\nu,\xi),
\label{eq:eb5.1}
\eeq
where $\xi_\nu=\xi(t_\non)$, $\Gamma_{\nu0}$ is evaluated at the initial density, $\rho_0$, and 
\beq
I_q(\xi_1,\xi_2)\equiv\frac{1}{\phiff\tffo}\int_{t(\xi_1)}^{t(\xi_2)} \xi(t')^q dt' 
\label{eq:iq00}
\eeq
is evaluated in Appendix \ref{app:free}, including the effects of dark matter. Here $\tffo$ is the free-fall time for the gas alone and $\phiff$ is a parameter of order unity that allows the collapse time for the gas alone to differ from $\tffo$ due to the fact that the collapse is not pressureless, for example. Observe that $dt\propto d\tff\propto d\xi/\xi^{3/2}$ so that $I_q$ is a number of order unity for $q<\frac 12$ and $\xi\gg1$.

Define the dynamo amplification factor $\cala(t)$ by
\beq
B(t)\equiv B_0\cala(t)\xi^{2/3};
\label{eq:bt}
\eeq
in terms of the specific magnetic energy, this is
\beq
\eb=\ebo\cala^2(t)\xi^{1/3}.
\label{eq:eb6}
\eeq
In the kinematic phase, equations (\ref{eq:eb3}) and (\ref{eq:eb3eta}) show that $\cala=\cala_\kin$ is exponentially sensitive to the input parameters. For the nonlinear phase,
we have 
\beq
B=B_\nu\cala_\non(\xi/\xi_\nu)^{2/3}~~~~~~(\xi>\xi_\nu),
\label{eq:banl}
\eeq
where
\beq
\cala_\non=\left[1+2\phiff\chi \Gamma_{\nu 0}\tffo\xi_\nu^{\frac 13+ q_\Gamma-q_\epsilon} I_{q_\epsilon-1/3}(\xi_\nu,\xi)
\right]^{1/2}
\label{eq:anl2}
\eeq
from equation (\ref{eq:eb5.1}) after expressing $\eb$ in terms of $B$.
Note that the second term is proportional to
\beq
\Gamma_{\nu0}\tffo=\left(\frac{\epsilon_0}{\nu_0}\right)^\frac 12 \tffo
=\left(\frac{v_t\tffo}{L_0}\right)Re^\frac 12.
\label{eq:gtf}
\eeq
For gravitational collapse, the factor in parentheses in the final expression is of order unity, so it follows that $\cala_\non\propto Re^{1/4}$ for large $Re$.

We now show that the nonlinear dynamo amplifies the field to a significant fraction of equipartition provided the dynamo amplification factor is large ($\cala_\non^2\gg1$). First consider the case in which the kinematic stage of the dynamo ends early in the collapse, so that $\xi_\nu\sim 1$. Since $\epsilon_0=v_{t0}^3/L_0$, equation (\ref{eq:eb5.1}) implies
\beq
\frac{\eb}{\frac 12 v_{t0}^2}\simeq 2\chi\left(\frac{\phiff v_{t0}\tffo}{L_0}\right) \xi^{1/3}I_{q_\epsilon-\frac 13}(1,\xi).
\label{eq:ebeq1}
\eeq
The factor in parentheses is of order unity; for example, for sonic turbulence in which the outer scale of the turbulence is the Jeans length, $v_{t0}\tffo/L_0=(3/32)^{1/2}$.
As noted above, when $q<\frac 12$, corresponding to $q_\epsilon<\frac 56$, the factor $I_q$ is a number of order unity for $\xi\gg 1$; on the other hand, for $q\geq \frac 12$, $I_q$ is an increasing function of $\xi$. It follows that even in the absence of the compression factor $\xi^{1/3}$, the nonlinear dynamo will bring the field up to an energy of order $2\chi\sim 0.1$ of equipartition. In the opposite case in which the nonlinear stage of the dynamo begins late in the collapse ($\xi_\nu\gg1$), $I_q(\xi_\nu,\xi)$ can be inferred from equation (\ref{eq:iq1}). As a result, the field energy for $\xi\gg\xi_\nu$ is
\beq
\frac{\eb}{\frac 12 v_{t\nu}^2}\simeq 2\chi\left(\frac{\phiff v_{t\nu}t_{\rm ff,\nu}}{L_\nu}\right)\frac{2}{3\pi(\frac 56-q_\epsilon)}\left(\frac{\xi}{\xi_\nu}\right)^{1/3}
\label{eq:ebeq2}
\eeq
for $q_\epsilon<\frac 56$, where $v_{t\nu}$ is the turbulent velocity at a density $\rho(t_\non)=\rho(\xi_\nu)$, etc. For $q_\epsilon\geq \frac 56$, the field energy is larger than this. Hence, for $\cala_\non^2\gg 1$, the nonlinear dynamo is efficient at bringing the field close to equipartition when $\xi_\nu\gg 1$ as well. In both cases, the relative importance of amplification of the field by the nonlinear dynamo and by compression  is
given by the ratio $\cala_\non(\xi_\nu/\xi)^{2/3}$. By contrast, this ratio for the specific magnetic energy, $\eb$, is $\cala_\non^2(\xi_\nu/\xi)^{1/3}$, which is generally much larger.

As remarked above, \citet{laza15} have argued that flux freezing is violated due to reconnection in a turbulent medium. We note that the effect of eliminating the effect of compression in the evolution of the nonlinear dynamo (i.e., omitting the second term in equation \ref{eq:deb4}) would be to omit the factors of $\xi$ and $\xi_\nu$ and replace $q_\epsilon-\frac 13$ by $q_\epsilon$ in equations (\ref{eq:ebeq1}) and (\ref{eq:ebeq2}); this would not affect the conclusion that the nonlinear dynamo is capable of bringing the field close to equipartition in a gravitational collapse.


We now estimate the magnitude of the field in the gas that forms the first stars.


\section{Predicted Magnetic Field in the Formation of the First Stars}
\label{sec:predthy}

We first discuss the initial Biermann field expected in a minihalo (or galaxy) and then the final value that results from the turbulent cascade. We show that the Biermann field is amplified rapidly in the kinematic stage of a small-scale dynamo, so that the density in this stage is approximately equal to the initial value. In the nonlinear phase of evolution of the dynamo, the field is amplified primarily by the compression due to the gravitational collapse that leads to star formation. This compression drives the field to equipartition, and it remains approximately in equipartition until non-ideal MHD effects take over. An overview of the predicted evolution of the field is shown in Fig. \ref{fig:thy}.

\begin{figure}
\hspace{-0.5cm}
\includegraphics[scale=0.45]{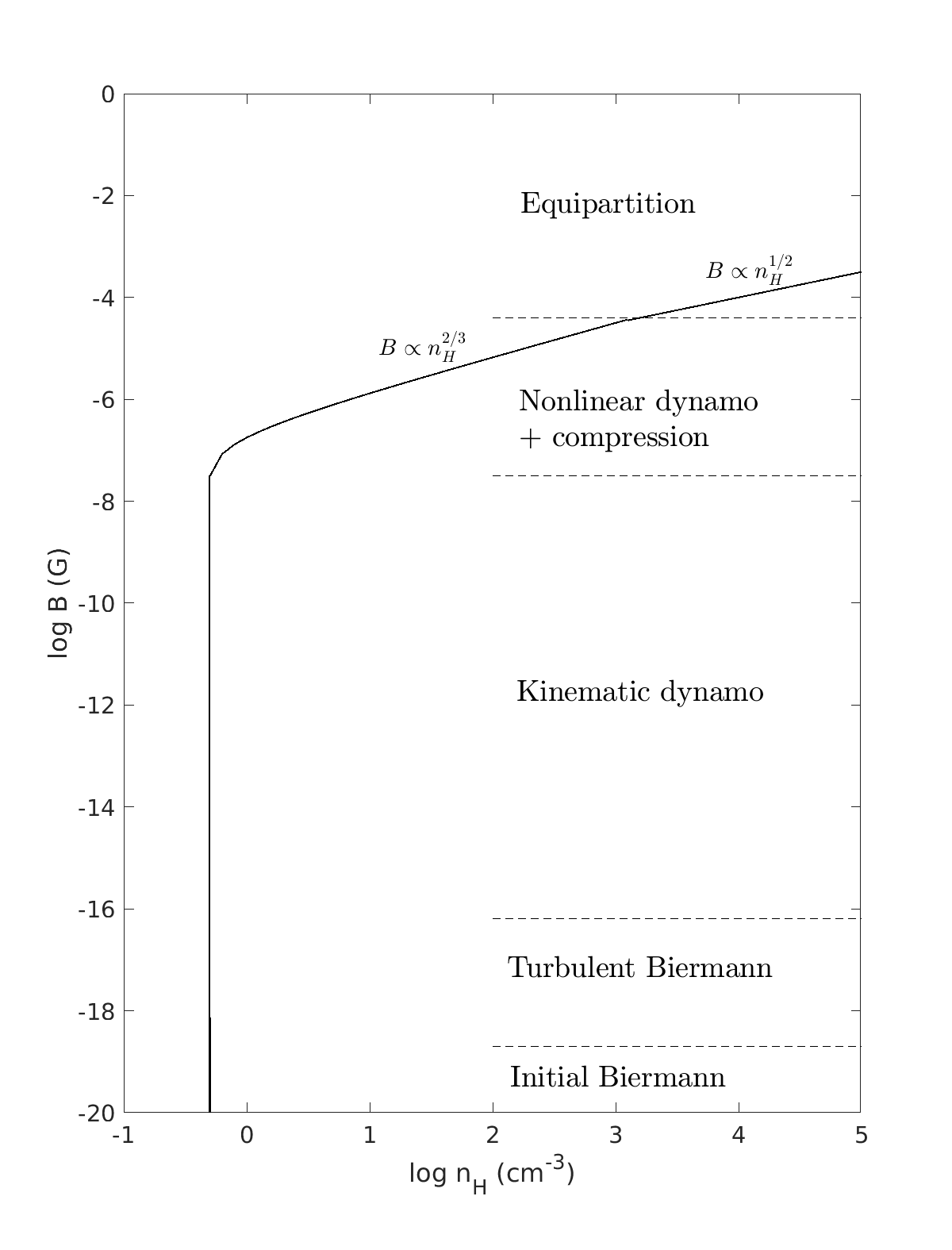}
\vspace{-1cm}
\caption{
\label{fig:thy}
Predicted growth of the magnetic field in a minihalo of total mass $3\times 10^5\,M_\odot$ at $z=25$ under the assumption that the initial field is zero. We have assumed that the turbulent velocity is half the virial velocity ($\phi_t=0.5$) and that $T=10^3$~K. The initial Biermann battery due to curved shocks on the scale of the minihalo generates a field of about $2\times 10^{-19}$~G. As turbulence cascades to smaller scales, the field generated by the Biermann battery increases to $\sim 10^{-16}$~G. A small-scale dynamo amplifies this field to about $3\times 10^{-8}$~G during the kinematic stage. Because the growth rate of the dynamo is so large in this stage, the density is about constant in time. The nonlinear dynamo is much slower, so the baryons are compressed by a factor $\sim 4000$ during this stage as they collapse due to the combined gravity of the dark matter and the baryons. In this stage, the field grows to about $5\times 10^{-5}$~G. The dynamo amplifies the field by only a factor $\sim 7$ in the nonlinear stage; most of the growth of the field is due to compression.
The nonlinear stage ends when the field reaches equipartition with the turbulence, with
$v_t\simeq 2$~km s\e. Subsequently, the field remains in approximate equipartition with the turbulence.
}
\end{figure}


\subsection{The Initial Field}
As discussed in the Introduction, processes in the very early universe might create comoving  fields in the range $B_c\sim 10^{-15}-10^{-12}$~G, 
but these processes are hypothetical. 
The Biermann battery process prior to reionization produces much weaker fields, $B\sim (10^{-25}-10^{-24})$~G, 
in the IGM \citep{naoz&narayan2013} or $B\sim 10^{-19}$~G in protogalaxies \citep{bier51}; the comoving fields are smaller by a factor $a^2=1/(1+z)^2$. However, these fields are based on well-established physics, so we focus on them here.

The field produced by the Biermann battery in a minihalo or a galaxy in the process of formation is due to the oblique shocks \citep{pudr89}
associated with the formation of these objects. As discussed in Section \ref{sec:bier}, the magnitude of this field is $1.29\times10^{-4}\omega$, where $\omega$ is the vorticity. 
We estimate the vorticity on the outer scale of the turbulence as $\omega\sim v_\vir/r_\vir$, where $r_\vir=(3M_m/4\pi\rho_{\rm mh})^{1/3}$ and $v_\vir=(GM_m/r_\vir)^{1/2}$ are the virial radius and velocity, respectively, where $M_m$ is the mass of all the matter in the halo, including the dark matter, and where $\rho_{\rm mh}$ is the average matter density in the minihalo. It follows that
\beq 
\omega\sim \frac{v_\vir}{r_\vir}=\left(\frac{4\pi G\rho_{\rm mh}}{3}\right)^{1/2}, 
\eeq 
where for a simple tophat model of the formation of the minihalo, $\rho_{\rm mh}$ is approximately $18\pi^2$ times the ambient density in the Hubble flow at that time (e.g., \citealp{barkana&loeb2001}),
\beq
\rho_{\rm mh}=\frac{27\pi H_0^2\Omega_m}{4G}\left(\frac{1+z}{26}\right)^3.
\eeq
Here we have normalized to a redshift of 25, since that is a typical redshift at which a minihalo collapses (\citealp{greifetal2012}, Stacy et al in preparation).  For simplicity, we henceforth make the approximation $1+z\simeq 26\ztf$, where $\ztf=z/25$, which 
is accurate to within 1\% for $20<z<33$ and accurate to 4\% for $z>12$. Following Stacy et al (in preparation), we set $H_0=70$~km~s\e~Mpc\e, $\Omega_m=0.30$ and $\Omega_b=0.04$. It follows that the matter density in the minihalo is
$\rho_{\rm mh}=8.62\times 10^{-24}\ztf^3$ g~cm\eee, so that $\omega\sim1.5\times 10^{-15}\ztf^{3/2}$~s\e\
and $B\sim 2.0\times 10^{-19}\ztf^{3/2}$~G at the outer scale of the turbulence. At $z\sim 25$, this field is almost exactly as \citet{bier51} estimated.

As discussed in Section \ref{sec:bier}, the turbulent cascade increases the vorticity, and therefore the field, on smaller scales. To evaluate the final Biermann field, which occurs on the viscous scale where the vorticity is a maximum ($\omega\simeq \Gamma_\nu$), and the properties of the subsequent dynamo,
we assume that the turbulence is governed by the properties of the minihalo. 
We then have for the outer scale of the turbulence in the minihalo $r\simeq r_\vir=
123M_{m,6}^{1/3}/\ztf$~pc,
where $M_{m,6}=M_m/(10^6\,M_\odot)$ (cf. \citealp{barkana&loeb2001}).
The virial velocity is $v_\vir=5.9 M_{m,6}^{1/3}\ztf^{1/2}$~km s\e.  
Simulations indicate that the turbulent velocity is somewhat less than this; for example,
the results of \citet{greifetal2012} show that $v_t\simeq 2$~km~s\e\ to within a factor 1.5 in the range
$r\sim 5-50$~pc for $M_m\simeq 3\times 10^5\,M_\odot$, corresponding to $v_t\simeq 0.5 v_\vir$, and a similar result was obtained by Stacy et al (in preparation). We therefore set
\beq
v_t=\phi_t v_\vir 
\eeq
and adopt $\phi_t=0.5$ as a fiducial value.
The density of hydrogen in the minihalo 
corresponding to the matter density $\rho_{\rm mh}$ is $\nh=(\Omega_b/\Omega_m)\rho_{\rm mh}/\muh=0.52\ztf^3$~cm\eee, where $\muh=2.23\times 10^{-24}$~g is the mass per H atom.
Equation (\ref{eq:turbbier}) then implies 
that the final Biermann field is
\beq
B_0=3.0\times 10^{-16}\left(\frac{\phi_t^{3/2}M_{m,6}^{1/3}
\ztf^{11/4}}{T_3^{0.42}}\right)~~~\mbox{G}.
\label{eq:bo}
\eeq
For a minihalo with $M_m=3\times 10^5\,M_\odot$ at $z=25$, this gives $B_0\simeq 7\times 10^{-17}$~G (with $\phi_t=0.5$ and $T_3=1$). 

\subsection{The Kinematic Dynamo}

The field produced by the Biermann battery is too weak to have any dynamical effects, so the dynamo begins in the kinematic, dissipation-free stage and the field exponentiates as $B\propto \exp{(\Gamma_\nu t)}$ (Section \ref{sec:ideal}). In order to determine the subsequent evolution of the field, we must first determine how long the kinematic stage lasts in comparison with the dynamical time of the minihalo. Ambipolar diffusion is the dominant dissipation mechanism for $B\ga 10^{-13}\nh$~G (Appendix \ref{app:ohm}), and
as discussed in Section \ref{sec:ambi}, the properties of the dynamo in the presence of ambipolar diffusion are governed by
dynamo ionization parameter, $\calr=6\nuni/\Gamma_\nu$ (equation \ref{eq:calr}). Using the just cited values of the density and radius of the minihalo, we find
\beq
\calr=1.2\left[\frac{\phi_d \xif T_3^{0.80}\ztf^{1/4}}{(2\phi_t)^{3/2} M_{m,6}^{1/3}}\right]
\label{eq:calr2}
\eeq
from Table \ref{tab:dyn}, where $\xif=x_i/10^{-4}$ is the normalized ionization fraction. The results of \citet{greifetal2012} give $\xif\sim 1$ for $r\ga 10$~pc. For $T\ga 500$~K, which is generally the case for the average gas in the minihalo \citep{abeletal2002,greifetal2012}, this implies $\calr\ga 0.7/M_{m,6}^{1/3}$. 
This is larger than the value found by \citet{xu&lazarian2016} since the ion-neutral collision rate in the post-recombination universe is larger than the value they adopted, as discussed in Appendix \ref{app:viscosity}.
Since $\calr$ is of order unity, the evolution of the dynamo is intermediate between the tracks shown in the bottom two parts of Fig. \ref{fig:AD}, so the scale of the turbulent field in the kinematic stage remains constant at about $\ell_\nu$.
Furthermore, we can use equation (\ref{eq:tnl2}) for the time at which the dynamo becomes nonlinear, $t_\non$.
Recall that $t_\non\propto\ln(B_\nu /(B_0\xi_\nu^{2/3})$ and that the initial field in the minihalo is $B_0\simeq 10^{-16}$~G from equation (\ref{eq:bo}).
Initially, the dynamics of the gas in the minihalo are determined by the dynamical time, $t_\vir=r_\vir/v_\vir=20.4\ztf^{-3/2}$~Myr.
Anticipating that $t_\non/t_\vir$ will be $< 1$, we infer that the density is about constant so that the density at the end of the kinematic stage, $\rho_\nu$, is about the same as the initial density (i.e., $\xi_\nu\simeq 1$) and 
$\avg{\Gamma_\nu}\simeq\Gamma_\nu$. We then obtain
\beq
B_\nu=5.80\times 10^{-8} \phi_t^{3/4}T_3^{0.21}\ztf^{11/8}M_{m,6}^{1/6}~~~\mbox{G}
\label{eq:bnu1.5}
\eeq
from Table \ref{tab:dyn}. In evaluating $t_\non$, we set $\phi_t=\frac 12$, and in the logarithmic factor we set the remaining parameters equal to unity, so that
\beq
\frac{t_{\non}}{t_\vir}\simeq 0.10\left(\frac{T_3^{0.42}}{M_{m,6}^{1/3}\ztf^{5/4}}\right).
\label{eq:tnonvir}
\eeq
We conclude that for a typical minihalo, the dynamo can reach a nonlinear amplitude in a time significantly less than the virial time. 

Reference to Fig. \ref{fig:AD} shows that the dissipation-free stage in the kinematic dynamo lasts until $\eb=(\calr^2\ebo/E_\nu)^{1/3}E_\nu$, which corresponds to a magnetic field $B\simeq (\calr^{1/2} B_\nu/B_0)^{2/3}B_0$. For $\calr$ and the remaining parameters all of order unity, this implies that the field is amplified by almost factor of $10^6$ before dissipation becomes important. Once that occurs, the field grows more slowly, $B\propto \exp(\Gamma_\nu t/6)$
\citep{kuls92,xu&lazarian2016}. For $\calr\ga 1$, as is the case here, this exponential growth continues until the dynamo reaches the nonlinear stage at $t=t_\non$.

\subsection{The Nonlinear Dynamo}

As noted above, the value of the dynamo ionization parameter, $\calr$, is initially of order unity. As the gas collapses in the nonlinear stage of the dynamo,
$\calr\propto x_i\nh^{1/2} r^{1/2}$ from Table 1. Since the gas is in ionization equilibrium, the ionization varies as $\nh^{-1/2}$ so that
$\calr\propto r^{1/2}\propto\xi^{-1/6}$. As we shall see, dynamo amplification in the nonlinear stage is significant only during the initial stages of the collapse, so we shall continue to use the results for for $\calr\ga 1$.
The field is then given by equation (\ref{eq:banl}) with $\cala_\non$ given by equation (\ref{eq:anl2}).
The nonlinear amplification factor $\cala_\non$ depends on how the energy dissipation rate depends on density, $\epsilon\propto\rho^{q_\epsilon}$, with $q_\epsilon=3q_v-q_L$,
through the factor $I_{q_\epsilon-1/3}(\xi_\nu,\xi)$ (equation \ref{eq:anl2}). 
Since $t_\non$ is only a fraction of the dynamical time, $t_\vir$, it follows that the density at $t_\non$ is close to the initial density,
$\rho_\nu\simeq \rho_0$,
so that $\xi(t_\non)=\xi_\nu\simeq 1$. The maximum value of $I_q(1,\xi)$ is reached when the collapse is complete, and as shown in equation (\ref{eq:iqapp}), it is of order unity provided that $q_\epsilon-\frac 13 < \frac 12$, which it generally is. Simulations such as those of \citet{greifetal2012} show that although the turbulent velocity is roughly constant, it does vary by a factor $\la 3$ in a complex manner, so the effective value of $q_\epsilon$ is uncertain. For a simple analytic estimate, we shall take advantage of the fact that $I_{q,\infty}=\calo(1)$ and set $q=q_\epsilon-\frac 13=0$. Equation (\ref{eq:iq00}) then gives $I_{0,\infty}=\tcoll/(\phiff\tffo)$. 
 Approximating the collapse time as $t_\vir$ and recalling that $t_\non\ll t_\vir$, we find from equation (\ref{eq:anl2})
that the total amplification by the nonlinear dynamo is
\beq
\cala_{\rm nl,\,tot}\sim\left(1+2\chi\Gamma_{\nu 0}t_\vir\right)^{1/2}.
\label{eq:anl3}
\eeq
Noting that $t_\vir=(3/4\pi G\rho_{\rm mh})^{1/2}$, we find
\beq
\cala_{\rm nl,\,tot}\sim\left[1+66(2\phi_t)^{3/2}\ztf^{5/4}T_3^{-0.42}M_{m,6}^{1/3}\right]^{1/2},
\label{eq:anl4}
\eeq
so the nonlinear dynamo amplifies the field by less than an order of magnitude in a minihalo. This relatively small amplification is because the field energy grows linearly in time in the nonlinear dynamo, but the time available for growth varies as $\xi^{-1/2}$ and is small in the late stages of the collapse. 
Using equations (\ref{eq:rho}) and (\ref{eq:iq1})}, one can show that 90\% of the amplification by the dynamo is completed before the time that $\xi=30$. (The fact that the dynamo amplification is concentrated in the early stages of the collapse justifies our assumption that we can follow the evolution of the nonlinear dynamo with the initial value of $\calr\propto \xi^{-1/6}$, which is of order unity.) As shown in Fig. \ref{fig:thy}, the growth of the field is dominated by compression ($B\propto \xi^{2/3}\propto \nh^{2/3}$) for most of the nonlinear stage.

\subsection{Equipartition}
\label{sec:equi}

As the collapse continues, 
the field eventually reaches approximate equipartition with the turbulence. 
When does this occur? We anticipate that it occurs only after significant compression, at a time close to the time $\tcoll$ at which the gas in the minihalo has collapsed. 
Now, for $B>B_\nu$ we have
\beq
\eb=\ebn\cala_\non^2\xi^{1/3}
\eeq
from equation (\ref{eq:banl}) with $\xi_\nu\simeq 1$. With the aid of equations (\ref{eq:ebn1}) and (\ref{eq:ellnu}), we have $\ebn\simeq \frac 12 (\epsilon_0\nu_0)^{1/2}=\frac 12\epsilon_0/\Gamma_{\nu 0}$. Since $\cala_{\rm nl,\,tot}^2\gg1$ from equation (\ref{eq:anl4}), it follows that the first term in 
equation (\ref{eq:anl3}), representing the field due to the kinematic dynamo, is negligible. We then have
\beq
\eb\simeq \chi\epsilon_0\xi^{1/3}t_\vir~~~~~(t\simeq t_{\rm coll}\simeq t_\vir).
\label{eq:eb}
\eeq
Since $\epsilon_0t_\vir\simeq (v_t^3/r_\vir)(r_\vir/v_\vir)$, this implies that
\beq
\frac{\va^2}{v_t^2}=\frac{\eb}{\frac 12 v_t^2}\simeq 2\chi\xi^{1/3}\frac{v_t}{v_\vir}=2\phi_t\chi\xi^{1/3}.
\label{eq:vavt}
\eeq
Equipartition first occurs when this ratio is unity, 
corresponding to a compression of
\beq
\xi_\eqi =\frac{1}{(2\phi_t\chi)^3}\simeq \frac{4100}{(2\phi_t)^3},
\label{eq:xieq}
\eeq
which is only a small fraction of the total compression the gas experiences as it collapses into a protostar.
Note that this condition for equipartition is independent of all the dimensional parameters of the problem, as expected from equations (\ref{eq:ebeq1}) and (\ref{eq:ebeq2}).
The corresponding density is
\beq
\nhei=\left(\frac{1}{2\phi_t \chi}\right)^3\nho \simeq 2.1\times 10^3\left(\frac{\ztf}{2\phi_t}\right)^3~~~\mbox{cm\eee}.
\label{eq:nhe}
\eeq
The initial equipartition magnetic field is then
\beq
B_\eqi=(4\pi\rho_0v_t^2)^{1/2}\xi_\eqi^{1/2}\simeq 7.2\times 10^{-5}\left[\frac{\ztf^2 M_{m,6}^{1/3}}{(2\phi_t)^{1/2}}\right]~~~\mbox{G}.
\label{eq:beqi}
\eeq

As noted by \citet{schleicheretal2010}, we expect that
once the field reaches equipartition, it will remain there as the compression continues, so that
the field will increase as $\rho^{1/2}$ (for a constant turbulent velocity) rather than $\rho^{2/3}$ (see Fig. \ref{fig:thy}).
This behavior is consistent with the results of the simulations of collapsing turbulent cores by \citet{mocz17}, who found that
the field remained close to equipartition with the turbulent energy as the density increased by orders of magnitude.
For an initially weak field, they found that the field eventually increased as $\rho^{2/3}$, presumably because the turbulent velocity increased near the nascent protostar; we note that if $v_t^2\propto r^{-1}$, then $B_\eq^2\propto \rho/r\propto \rho^{4/3}$.
Our conclusion that the dynamo reaches equipartition in the formation of stars at $z\sim 25$ differs from that
of \citet{xu&lazarian2016}, who concluded that equipartition is reached at $t\sim 6\times 10^8$ yr (corresponding to $z\sim 8$), because
they did not consider the increase in density that accompanies star formation.

As noted in Section \ref{sec:ideal}, it is possible that the field could saturate at a value different than the equipartition value, 
\beq
B_\rms^2=B_\sat^2=4\pi\phi_\sat^2\rho v_t^2,
\label{eq:sat}
\eeq
with $\phi_\sat$ most likely somewhat less than 1.
In that case, the \alfven\ Mach number in the saturated state would be
$\ma=v_t/\va=1/\phi_\sat$; the field would be dynamically insignificant for $\phi_\sat\ll 1$. 
Equation (\ref{eq:vavt}) implies that the field saturates at a compression
$\xi_\sat=\phi_\sat^6\xi_\eqi$. For $\phi_\sat=0.7$, corresponding to the subsonic turbulence (e.g., \citealp{federrath11a}) relevant for the formation of the first stars \citep{abeletal2002,greifetal2012}, this gives
$\xi_\sat=480/(2\phi_t)^3$.

\subsection{The Magnetic Field vs Gravity}
\label{sec:gravity}

How does the force associated with the magnetic field compare with that due to gravity? The magnetic critical mass
is the mass for which the gravitational and magnetic forces balance. There are two forms for the critical mass, $M_\Phi\simeq \Phi/(2\pi G^{1/2})$, where $\Phi=\pi r^2 B_\rms$ is the magnetic flux based on the rms field in the
cloud, and 
\beq
M_B=\frac{M_\Phi^3}{M^2}=\frac{9}{128\pi^2 G^{3/2}}\left(\frac{{B_\rms}^2}{{\rho}^{4/3}}\right)^{3/2},
\label{eq:mb0}
\eeq
where $M(r)$ is the gas mass inside $r$ (e.g., \citealp{mckee&ostriker2007}).
The force of gravity exceeds that due to magnetic fields for
$M>M_\Phi$ or $M>M_B$, so a necessary condition for gravitational collapse is that these inequalities be satisfied
(note that $M_\Phi=M_B$ for $M=M_\Phi$, so that this is actually a single condition). 

As the baryons collapse, they form a core with a power-law density profile, $\rho\propto r^{-\krho}$ with $\krho\simeq 2.2$.
For example, a fit to the results of \citet{greifetal2012} and Stacy et al (in preparation) give $\krho\simeq 2.3$ and 2.16, respectively, while the theoretical model of \citet{tan04} has $\krho=20/9\simeq 2.22$. 
The fraction of the mass with a density greater than $\rho$ is then 
\beq
M(>\rho)/M_0=M(>\xi)/M_0=\xi^{-(3-\krho)/\krho},
\label{eq:krho}
\eeq
where $M_0$ is the total mass of gas in minihalo; for $\krho=2.2$, this is
$M(>\xi)/M_0=\xi^{-0.36}$. The field is at its equipartition value for the inner 5\% of the core for $\phi_t=\frac 12$ since
$M(>\xi_\eq)/M_0=0.05(2\phi_t)^{1.08}$. As an example, for a minihalo of mass $3\times 10^5\,M_\odot$, we have $M_0=4\times 10^4\,M_\odot$, so that the central $2000\,M_\odot$ has an equipartition field.
If the field saturates at a value other than the equipartition value, then the mass of gas with a saturated field would be $2000\phi_\sat^{-2.16}\,M_\odot$, which is $4300\,M_\odot$ for $\phi_\sat=0.7$.

We have seen that most of the amplification of the field in the nonlinear stage is due to compression, so that $B_\rms$ scales approximately as $\rho^{2/3}$ prior to equipartition ($\xi<\xi_\eq$); it follows that $M_B$ is approximately constant during this phase. Under the assumption that the turbulent velocity remains about constant, after equipartition we have $B_\rms^2=4\pi\rho v_t^2$ so that $M_B\propto\rho^{-1/2}$. To cover both cases, we note that equations (\ref{eq:vavt}) and (\ref{eq:xieq}) imply
\beq
B_\rms^2=\min\left[\left(\frac{\xi}{\xi_\eq}\right)^{\frac 13},1\right]4\pi\rho v_t^2
\label{eq:brms}
\eeq
for pre- and post-equipartition, respectively. From equation (\ref{eq:mb0}) 
we then find that magnetic fields limit the mass that can undergo gravitational collapse to be at least
\beqa
M_B&\hspace{-0.3cm}=\hspace{-0.3cm}&
\left[\frac{3\chi(2\phi_t)^3}{16}\right]^{\frac 32}\left(\frac{\Omega_m}{\Omega_b}\right)^{\frac 12}\min\left[1,\left(\frac{\xi_\eqi}{\xi}\right)^{\frac 12}\right]M_m,~~~~~~
\label{eq:mb1}\\
&\hspace{-0.3cm}=\hspace{-0.3cm}&3470\left(2\phi_t\right)^3\min\left[(2\phi_t)^{\frac 32},\left(\frac{4100}{\xi}\right)^{\frac 12}\right]M_{m,6}~~M_\odot,
\label{eq:mb2}
\eeqa
where we set $\chi=1/16$ and $\Omega_m=7.5\Omega_b$ in the second equation. 
Note that equation (\ref{eq:mb1}) applies to present-day GMCs for equipartition fields if $\xi_\eq$ is inserted from equation (\ref{eq:xieq}) and $\Omega_m$ is set to $\Omega_b$.
Since $\phi_t=v_t/v_\vir$, the value of $M_B$ is very sensitive to the turbulent velocity, $v_t$.
Prior to equipartition (first term in the above equations), $M_B$ is constant, but for $\xi>\xi_\eq$ (second term), $M_B$ varies as $\xi^{-1/2}\propto r^{\krho/2}$.
In order for gravity to overcome magnetic fields for masses much less than $3500\,M_\odot$, high densities are required; 
for example, reducing $M_B$ below $100\,M_\odot$ requires $\nh\ga 10^{6.5}$~cm\eee\ for $\phi_t\sim \frac 12$ and $M_{m,6}\sim\ztf\sim 1$. 

To compare with contemporary star formation, we recast these results in terms of the ratio of the gas mass inside $r$ to the critical mass at that radius, 
\beq
\mu_\Phi\equiv \frac{M(r)}{M_\Phi(r)}=\left[\frac{M(r)}{M_B}\right]^{\frac 13}.
\eeq
Equation (\ref{eq:mb1}) then implies that
\beq
\mu_\Phi=\frac{4\xi^{(\krho-2)/2\krho}}{(2\phi_t)\surd 3} \left(\frac{\Omega_b}{\Omega_m}\right)^{\frac 12}\max\left[\left(\frac{\xi_\eqi}{\xi}\right)^{\frac 16},1\right].
\label{eq:muphi1}
\eeq
Just as in the case of equation (\ref{eq:mb1}) for $M_B$, this result applies to GMCs for equipartition fields if $\xi_\eq$ is inserted from equation (\ref{eq:xieq}) and $\Omega_m$ is set to $\Omega_b$. For the particular case $\krho=2.2$ and $\phi_t=\frac 12$, equation (\ref{eq:muphi1}) becomes
\beq
\mu_\Phi=1.23\max\left[\left(\frac{4100}{\xi}\right)^{0.12},\left(\frac{\xi}{4100}\right)^{0.045}\right].
\label{eq:muphi2}
\eeq
Note that the density dependence of $\mu_\Phi$ is weak: The entire minihalo ($\xi=1$) has $\mu_\Phi=3.4$; the minimum value, $\mu_\Phi=1.23$, occurs at the point that the gas first reaches equipartition ($\xi=4100$); and $\xi$ must exceed $2\times 10^{13}$ in order for $\mu_\Phi$ to exceed 3.4. Over this entire density range, $\mu_\Phi\simeq 2\pm 0.2$~dex.

As noted above, the field might saturate at a value that differs from the equipartition value by a factor $\phi_\sat$, and
correspondingly, 
$\mu_\Phi$ would differ from the values given in equations (\ref{eq:muphi1}) 
and (\ref{eq:muphi2}) by a factor $1/\phi_\sat$. The Mach number in the simulations of \citet{abeletal2002} is of order 1/3, which is subsonic, so that
$\phi_\sat\sim 0.7$ \citep{haug04,federrath11a} and $\mu_\Phi\sim 2/0.7\sim 3$; the simulations of \citet{greifetal2012} have Mach numbers $\sim 1$, which would give a somewhat larger value of $\mu_\Phi$.

The results we have obtained for the magnetic fields in a minihalo are quite comparable to those for the fields in contemporary star-forming regions. Equation (\ref{eq:nhe}) shows that the field is in equipartition with turbulent motions at densities $\ga 10^3$ cm\eee, comparable to the densities in molecular clumps today. As discussed above,
equation (\ref{eq:muphi2}) shows that the equipartition value of the mass-to-flux ratio is $\mu_\Phi\sim 2$, which is the value expected on theoretical grounds for Galactic GMCs \citep{mckee89};
at present, there is no direct measurement available for $\mu_\Phi$ for GMCs. 
Star-forming clumps within GMCs have $\mu_\Phi\simeq 2-3$ \citep{crutcher12,li15}, which is also in good agreement with the predicted value in equation (\ref{eq:muphi2}).  

\citet{krum19} have recently reviewed the role of magnetic fields in 
contemporary star formation. For typical mass-to-flux ratios ($\mu_\Phi\sim 2-3$), magnetic fields reduce the rate of star formation by a factor of a few. Magnetic fields have little direct effect on the peak of the IMF since radiative feedback is generally dominant. Magnetic fields reduce fragmentation, particularly in disks, which could suppress the formation of low-mass primordial stars that could survive until today. Reduced fragmentation also favors the production of massive stars. One of the main effects of magnetic fields is that if they are ordered, they produce outflows that reduce the typical stellar mass by a factor $\sim 2-3$. However, recent simulations show that no outflows are produced by turbulent magnetic fields \citep{gerr19}, so that effect should not be present in primordial star formation.

In sum, the kinematic dynamo is able to amplify the field from very small values ($\sim 10^{-25}-10^{-19}$~G) to moderate values $\sim 10^{-8}$~G, with very little of the amplification due to compression. On the other hand, the nonlinear dynamo is much less efficient, providing an amplification of less than an order of magnitude in our example. The initial equipartition field $\sim 10^{-4}$~G is attained with a compression somewhat less than $10^4$, and we anticipate that the field will remain in approximate equipartition as the collapse continues to higher densities. During this phase of the collapse, the mass supported by the field against gravity, $M_B$, declines as $\xi^{-1/2}$ (equation \ref{eq:mb2}) so that the mass-to-flux ratio in the core is nearly independent of density (equation \ref{eq:muphi2}). The equipartition field, as characterized by the ratio of the turbulent velocity to the virial velocity,
$\phi_t=v_t/v_\vir\sim \frac 12$, results in a normalized mass-to-flux ratio, $\mu_\Phi$, somewhat above unity.  We estimate $\mu_\Phi \sim 3$ for subsonic turbulence, comparable to that in contemporary star-forming regions. As a result
magnetic fields could play a role in the formation of the first stars.

\section{Theory of Simulations}
\label{sec:predsim}

One of the principal difficulties in simulating astrophysical fluids is that the physical viscosity is generally orders of magnitude smaller than the numerical viscosity,
so that the actual Reynolds number is orders of magnitude larger than that in the simulation. For dynamos in mini-halos, the physical viscosity is set by collisions in neutral hydrogen and is $\nu\sim 10^{20}$~cm$^2$~s\e\ for $T_3\sim 1$ and $\nh\sim 1$~cm\eee\ (Appendix \ref{app:viscosity}), whereas the numerical viscosity in SPH or grid-based codes is of order $10^{23}$~cm$^2$~s\e\ for the same physical conditions and
for resolutions corresponding to about 64 cells per Jeans length. 
As a result, the characteristic
growth rate in the kinematic stage of the dynamo, $\sim\Gamma_\nu\propto\nu^{-1/2}$ (equation \ref{eq:gamma}), is smaller by
a factor $\sim 10^{1.5}$. A corollary of this is that the time at which the dynamo enters the nonlinear stage, $t_\non\propto
\Gamma_\nu^{-1}$ (equation \ref{eq:tnl2}), is larger by about the same factor. Thus, whereas the actual mini-halo dynamo enters the nonlinear stage prior to significant compression, simulated mini-halo dynamos do so only after significant compression. We must
therefore use the results for a dynamo in a time-dependent background given in Section \ref{sec:time}. 

Another important difference between the simulations considered here and reality is that we assume that the simulations are based on ideal MHD, so that the resistivity is numerical. As a result, the resistivity in the simulations is independent of $B$, whereas in the weakly ionized plasma that forms the first stars it is dominated by ambipolar diffusion and varies as $B^2$; the effect of this approximation is less significant than the large discrepancy between the simulated and actual viscosities, however. 

The theoretically predicted evolution of the magnetic field shown in Fig. \ref{fig:ohm} is dramatically different from that in the simulations of \citet{turketal2012} and Stacy et al (in preparation), principally due to the difference between the actual viscosity and that in the simulations. As noted by \citet{suretal2010} and \citet{turketal2012}, the growth rate of the dynamo increases with the Reynolds number and therefore with resolution. (This follows directly from the growth rate of the kinematic dynamo $\Gamma\sim\Gamma_\nu$, (equation \ref{eq:gamma}), and the fact that $\Gamma_\nu\propto Re^{1/2}$ (eqs. \ref{eq:ellnu} and \ref{eq:re}).) Here we seek to predict the outcome of a simulation of the evolution of the magnetic field in the formation of the first stars so that we can understand how it relates to the theoretical expectation described in the previous section and portrayed in Fig. \ref{fig:thy}.

\subsection{SPH Simulations of Mini-halo Dynamos}
\label{sec:predsph}

We now estimate the outcome of an SPH simulation of a mini-halo dynamo. The numerical viscosity 
for SPH is
\beq
\nu_\sph=1.50\times 10^{23}\left(\frac{h_f m_\sph'^{1/3} T_3^{1/2}}{\nh^{1/3}}\right)~~~~~\mbox{cm$^2$ s\e},
\label{eq:nusph0}
\eeq
(equation \ref{eq:nusph2}),
where $h_f$ normalizes the SPH smoothing length (equation \ref{eq:hsm}) and
$m_\sph'=m_\sph/(1\,M_\odot)$ is the normalized SPH particle mass. For example, \citet{price2012b} adopted $h_f=1.2$, whereas Stacy et al (in preparation) adopted $h_f=3.63$; \cite{price2012b} did not need to adopt a value for $m_\sph$, but Stacy et al (in preparation) adopted $m_\sph\simeq 0.03\,M_\odot$ in the high resolution portion of their run, 
corresponding to $h_f m_\sph'^{1/3}=1.13$; their simulation had about $3\times 10^7$ particles representing the gas.
As noted above,
we expect the kinematic stage to extend well
into the gravitational collapse of the star forming in the mini-halo, and as a result the effective outer scale of the turbulence is the 
Jeans length \citep{federrath11b}, $\lj=386(T_3/\nh)^{1/2}$~pc.
The Reynolds number in the simulation of a gravitationally collapsing cloud is then
\beq
Re=\frac{\lj v_t}{\nu_\sph}=800\left(\frac{\vtf}{h_f m_\sph'^{1/3}\nh^{1/6}}\right).
\eeq

In order for a dynamo to operate, the magnetic Reynolds number, $R_m=P_m Re$, 
must exceed a critical value, $\rmcr$, as discussed in Section \ref{sec:resistive}. We adopt the result of \citet{haug04}, $\rmcr\simeq 220P_m^{-1/2}$ for $0.1\la P_m\la 3$, so that
\beq
\frac{R_m}{\rmcr}=3.6\left(\frac{P_m^{3/2}\vtf}{h_f m_\sph'^{1/3} \nh^{1/6}}\right).
\label{eq:rmrmcr1}
\eeq
The maximum density for the operation of the dynamo is determined by setting this ratio equal to unity,
\beq
n_{\rm H,\,max}=2.18\times 10^3\left(\frac{P_m^{3/2}\vtf}{h_f m_s'^{1/3}}\right)^6~~~\mbox{cm\eee}.
\label{eq:nhmax}
\eeq
In Appendix \ref{app:numres}, we estimate that the magnetic Prandtl number for grid-based codes is $P_m\simeq 1.4$, and we adopt the same value for SPH codes.  Then, for a typical turbulent velocity of 2~km~s\e\ \citep{greifetal2011a}, we
find that the dynamo can operate only below a density of $n_{\rm H,\,max}\simeq 3\times 10^6/(h_f m_\sph'^{1/3})^6$~cm\eee. This is in the upper range of the densities in the SPH simulation of
Stacy et al (in preparation), which has $h_f m_\sph'^{1/3}\simeq 1$.


\begin{figure*}
\includegraphics[width=\textwidth]{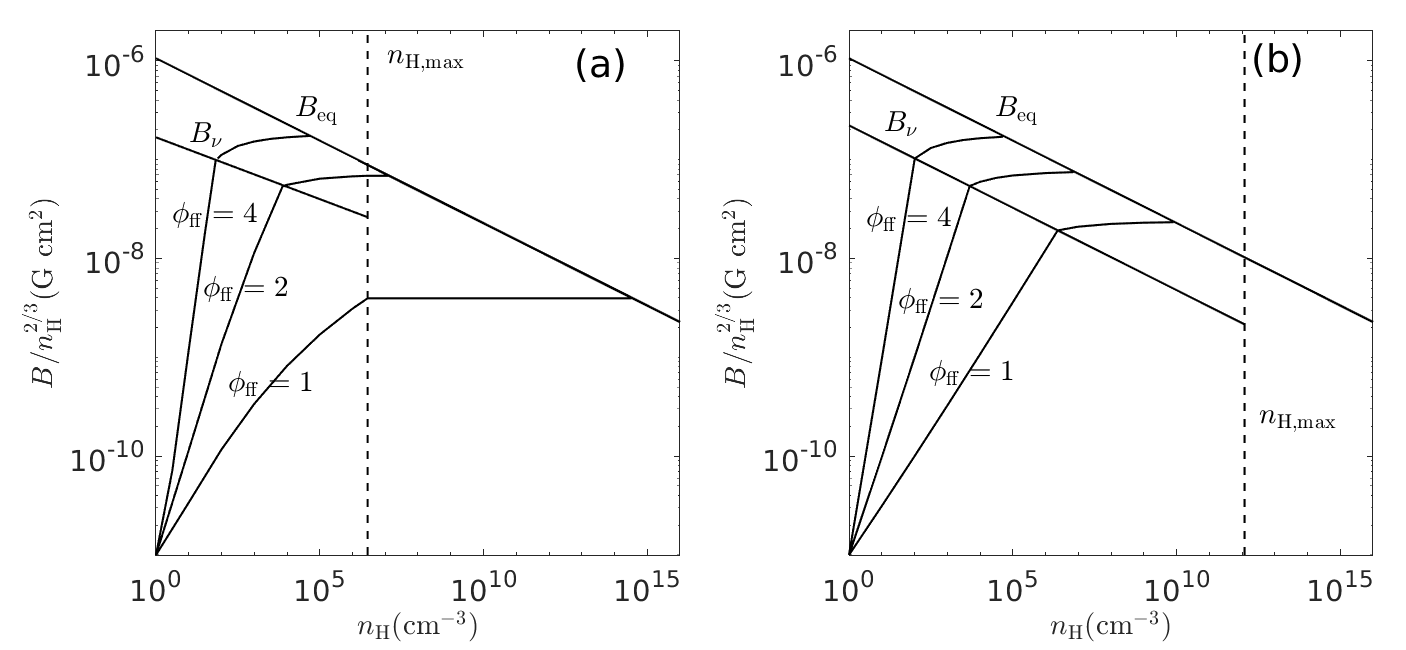}
\caption{
\label{fig:sim}
The expected value of $B/\nh^{2/3}$ for SPH (a) and grid-based (b) simulations of the gravitational collapse of a turbulent gas in a dark matter minihalo. Illustrated are cases in which the collapse  occurs at the free-fall rate ($\phiff=1$), half that rate $(\phiff=2$), and a quarter of that rate ($\phiff=4$). The equipartition field, $B_\eq=(4\pi\rho v_t^2)^{1/2}$ and the field at which nonlinear effects become important at the viscous scale, $B_\nu=(4\pi\rho v_\nu^2)^{1/2}$, both normalized by $\nh^{2/3}$, are also plotted. The initial field and density are $B_0=10^{-11}$~G and $\nho=1$~cm\eee, and we adopt $T=10^3$~K and $v_t=2$~km~s\e.  The magnetic Prandtl number for simulations is taken to be $P_m=1.4$ (Appendix \ref{app:numerical}.)
In reality (Fig. \ref{fig:thy}), the magnetic field 
begins at a value well below the minimum in this figure, intersects the nonlinear curve ($B_\nu$) at $\nh\simeq 1$~cm\eee, and then bends over to become nearly horizontal before intersecting the equipartition line ($B_\eq$). (a) The SPH simulation is assumed to have $h_f m_s'^{1/3}=1$ (see Appendix \ref{app:sphcodes}). For $\phiff=1$, the kinematic dynamo amplifies the field until the magnetic Reynolds number drops to the critical value below which the dynamo ceases, which occurs at a density $n_{\rm H,\,max}$. For $\phiff=2$ or 4, the field becomes nonlinear on the viscous scale ($B=B_\nu$). In any case, the nonlinear dynamo does not have much time to operate, so the field then grows primarily by compression in the nonlinear stage until it reaches equipartition with the turbulence. Subsequently, the field remains in approximate equipartition. (b)  As in the SPH case, the nonlinear dynamo does not have much time to operate, so the growth of the field in this stage is primarily by compression. The maximum density at which the dynamo can operate based on the condition $R_m<\rmcr$ is $\nhm=1.2\times 10^{12}$~cm\eee\ for the resolution of the AMR simulation of Stacy et al (in preparation).
}
\end{figure*}

In the kinematic phase of a simulated dynamo, the dynamo amplification factor is
\beq
\cala_\kin=\exp\left(\frac 38 \int_{t_0}^t\Gamma_\nu dt'\right)
\label{eq:akin}
\eeq
from equations (\ref{eq:eb3}) and (\ref{eq:eb6}). Here we have taken $k_p\ell_\nu\simeq 1$ in equation (\ref{eq:eb3}) since it lies between 1 and $P_m^{1/2}\simeq 1$ (see the middle panel of Fig. \ref{fig:ohm}).
Since our focus is on dynamos in gravitationally collapsing clouds, we consider
the case in which the growth rate varies as a power of the density,  $\Gamma_\nu=\Gamma_{\nu0}\xi^{q_\Gamma}$,
where $\xi=\rho/\rho_0$ is the compression ratio and, in general, $x\propto \xi^{q_x}$.
Recall that $\Gamma_\nu=(\epsilon/\nu)^{1/2}$ (equation \ref{eq:ellnu}) and $\epsilon=v_t^3/L$, so that if the outer scale of the turbulence is set by the Jeans length, then 
we have 
\beq
q_\Gamma=\frac 12 (q_\epsilon-q_\nu)
=\frac 12\left(3q_v-\frac 12 q_T+\frac 12-q_\nu\right).
\label{eq:qg}
\eeq
Simulations (e.g., \citealp{greifetal2011a}) show that whereas there is some variation of $v_t$ and $T$ in the collapse, it is not systematic, so we shall generally treat them as constant and set $q_v=q_T=0$. It follows that for SPH, $q_\nu=-\frac 13$ (equation \ref{eq:nusph0}), so that $q_\epsilon=\frac 12$ and $q_\Gamma=\frac{5}{12}$.

In Appendix \ref{app:free} we discuss the gravitational collapse of gas embedded in stationary dark matter.
We consider the idealized case in which both the gas and the dark matter initially have spatially constant densities so that the density of the gas remains spatially constant when it undergoes free-fall collapse. We assume that the infall velocity is a factor $\phiff$ below the free-fall value so that the collapse time is $\phiff$ times greater.
where $\tffo=(3\pi/32G\rho_0)^{1/2}$ is the initial free-fall time in the absence of dark matter.
The integral that appears in the dynamo amplification factor can be expressed as
\beq
\int_{t_0}^t\Gamma_\nu dt'=\Gamma_{\nu 0}\phiff\tffo I_{q_\Gamma}(1,\xi)
\label{eq:intg}
\eeq
in terms of the integral $I_q$ evaluated in Appendix \ref{app:free}; here the density dependence of $\Gamma_\nu$ is given by $\Gamma_\nu\propto \xi^{q_\Gamma}$.
Since the outer scale of the turbulence in a collapsing cloud is the Jeans length
\citep{federrath11b}, it follows that the factor $\epsilon$ that enters $\Gamma_\nu$ is
\beq
\epsilon=\frac{v_t^3}{\lj}=8.40\times 10^{-7}\vtf^3\left(\frac{\nh}{T_3}\right)^{1/2}~~~~~\mbox{cm$^2$ s\eee}.
\eeq
For the SPH viscosity given in equation (\ref{eq:nusph2}), we then have
\beq
\Gamma_{\nu0}\tffo=3.33\left(\frac{\vtf^{3/2}}{h_f^{1/2} m_\sph'^{1/6}T_3^{1/2}\nho^{\frac 12 -q_\Gamma}}\right),
\label{eq:gnu1}
\eeq
so that
\beq
\cala_\kin=\exp\left[ 1.25\left(\frac{\phiff \vtf^{3/2}}{h_f^{1/2} m_\sph'^{1/6} T_3^{1/2}\nho^{\frac 12 -q_\Gamma}}\right)I_{q_\Gamma}(1,\xi)\right]
\label{eq:akin2}
\eeq
from equations (\ref{eq:akin}) and (\ref{eq:intg}).

The growth of the field in a contracting medium is often characterized by the logarithmic derivative, $d\ln B/d\ln\rho$. For the kinematic stage of the dynamo, the field is
$B=B_0(\rho/\rho_0)^{2/3}\cala_\kin$ (equation \ref{eq:bt}). Since $\cala_\kin$ is given by equation (\ref{eq:akin}), we have
\beqa
\frac{d\ln B}{d\ln\rho}&\hspace{-0.3cm}=\hspace{-0.3cm}&\frac 23 +\frac{d\ln\cala_\kin}{dt}\left(\frac{dt}{d\ln\rho}\right),
\label{eq:dlnb1}\\
&\hspace{-0.3cm}=\hspace{-0.3cm}&\frac 23 + \frac 38 \Gamma_\nu\left(\frac{r}{3|v|}\right),
\label{eq:dlnb2}
\eeqa
where in the last step we used $\rho\propto r^{-3}$.
Late in the collapse ($r\ll r_0$, $\xi^{1/3}\gg 1$), the velocity is $|v|\simeq v_g(r_0/r)^{1/2}=v_g\xi^{1/6}$ 
with $v_g\sim r_0/\tffo$ (see Eqs. \ref{eq:v2} and \ref{eq:vg}), so that
\beqa
\frac{d\ln B}{d\ln\rho}&\hspace{-0.3cm}=\hspace{-0.3cm}&\frac 23 + \frac{1}{4\pi}\,\phiff \Gamma_{\nu0}\tffo\xi^{q_\Gamma-\frac 12},
\label{eq:dlnb3}\\
&\hspace{-0.3cm}=\hspace{-0.3cm}&\frac 23 + 0.26\left(\frac{\phiff \vtf^{3/2}}{h_f^{1/2} m_\sph'^{1/6}T_3^{1/2}\nho^{1/12}}\right)\xi^{q_\Gamma-\frac 12}.
\label{eq:dlnb4}
\eeqa
So long as $q_\Gamma<\frac 12$, the variation of $B$ with $\rho$ in the kinematic stage approaches $B\propto\rho^{2/3}$ at high densities--i.e., it is compression, not the dynamo, that amplifies the field then. As we shall see below, the slope is driven to 2/3 when the dynamo leaves the kinematic stage.

As an example, consider the case in which
$v_t$ and $T$ do not have a systematic variation during the collapse (i.e., $q_v=q_T=0$). As noted above equation (\ref{eq:qg}), it follows that
$q_\Gamma=5/12$, so that
equation (\ref{eq:iq1}) gives
\beq
I_{5/12}(1,\xi)\simeq 2.43(1-1.05\xi^{-\frac{1}{12}})~~~~~(\xi^{\frac 13}\gg1),
\eeq
where we have evaluated $I_{5/12}(1,\infty)$ numerically.
For $\vtf\sim 2$, $\nho\sim 1$~cm\eee, and $T_3\sim 1$, we then find
\beq
\cala_\kin\simeq\exp\left[\frac{8.6\phiff}{h_f^{\frac 12}m_\sph'^{\frac 16}}\left(1-1.05\xi^{-\frac{1}{12}}\right)\right]~~~~(\xi^{\frac{1}{3}}\gg1).
\label{eq:akin3}
\eeq
The quantity $B/\nh^{2/3}$, which is just $B_0\cala_\kin$ in the kinematic stage, is plotted in Fig. \ref{fig:sim}a for three values of $\phiff$, providing a graphic demonstration of the exponential sensitivity of the simulated dynamo to the input parameters. Note that for a kinematic dynamo, an increase in resolution at a fixed value of $\xi$ (which is numerically the same as $\nh$ in Fig. \ref{fig:sim} since $\nho=1$~cm\eee\ there) is equivalent to an increase in $\phiff$; for example, increasing the linear resolution by a factor 2 corresponds to reducing $m_\sph$ by a factor 8 and increasing $\phiff$ by $\surd 2$.

First, consider the case in which $\phiff=1$, so that the collapse occurs at the free-fall rate. This is sufficiently rapid that the dynamo cannot reach the nonlinear stage before dynamo action is terminated because the density reaches $\nhm$ and $R_m$ drops below the critical value. 
In this example, and for $h_fm_s'^{1/3}\simeq 1$, equation (\ref{eq:akin3}) gives an amplification factor for the kinematic dynamo of $\cala_\kin(\xi_{\max})\sim 10^2.6$, where $\xi_{\max}=\nhm/\nho\simeq 3\times 10^6$. The growth of the field by compression ($\xi_{\max}^{2/3}\simeq 10^{4.3}$) is much greater than the growth due to the dynamo ($\sim 10^2.6$). The slope of $B(\rho)$ approaches $\frac 23$ in the kinematic stage, and is then driven to $\frac 23$ when the kinematic stage terminates.
For $\nh>\nhm$ the field grows by compression until it reaches equipartition. As shown in Fig. \ref{fig:sim}a, which is based on the assumption that the initial field is $B_0=10^{-11}$~G, this occurs at a density 
$sim 10^{15}$~cm\eee, corresponding to $M/M_0\sim 4\times10^{-6}$ for a power-law density profile with $\krho=2.2$ (equation \ref{eq:krho}). For a minihalo with a gas mass of $4\times 10^4\,M_\odot$, the mass that reaches equipartition is very small, $\sim 0.2\,M_\odot$. Thus, in this case, the magnetic field has a negligible effect throughout most of the core, at least up to the time that the protostar begins to form. If the initial field were less than $10^{-11}$~G, the magnetic field would be even less important.


Next consider the case $\phiff=2$, in which the collapse occurs at half the free-fall rate so that the dynamo has more time to act. In this case, the field grows to $B_\nu$ before the density reaches $\nhm$.
At this point the \alfven\ velocity $\va$ equals
the velocity of the viscous scale eddies, $v_\nu=(\epsilon\nu)^{1/4}$ (equation \ref{eq:ellnu}), so that
\beq
B_\nu=(4\pi\rho)^{\frac 12}v_\nu=1.00\times 10^{-7}\left(h_f m_s'^{\frac 13}\right)^{\frac 14}\vtf^{\frac 34}\nh^{\frac{13}{24}}~~~~\mbox{G},
\label{eq:bnu2}
\eeq
which is plotted in Fig. \ref{fig:sim}a. 
Since $\va=\vao\cala_\kin\xi^{1/6}$
up to the point that $B$ reaches $B_\nu$ (equation \ref{eq:bt}), the compression required for the field to reach $B_\nu$ is
\beq
\xi_\nu=\left(\frac{v_\nu}{\vao\cala_\kin}\right)^6=
\left(\frac{v_{\nu 0}}{\vao\cala_\kin}\right)^{12/[2-3(q_\epsilon+q_\nu)]}
\label{eq:xinu}
\eeq
where we used $v_\nu=v_{\nu 0}\xi^{(q_\epsilon+q_\nu)/4}$ in the second expression.
The exponential dependence on the uncertain parameters in $\cala_\kin$ that describe the collapse  (see equation \ref{eq:akin3}) means that $\xi_\nu$ is essentially unpredictable for simulations with a numerical viscosity
several orders of magnitude larger than the actual one, as is generally the case. By contrast, $\xi_\nu$ is well determined
in Nature: the small viscosity means that the exponent in the expression for $\cala_\kin$ (equation \ref{eq:akin2}) is large enough to make $\xi_\nu\simeq 1$ (Section \ref{sec:predthy}). 
For the hypothetical simulation with $\phiff=2$ shown in Fig. \ref{fig:sim}a, the field reaches $B_\nu$ at $\xi_\nu\simeq 10^4$ with $\cala_\kin\simeq 10^{3.7}$, so that the dynamo amplification is an order of magnitude greater than that due to compression. On the other hand, for $\phiff=4$ the field reaches $B_\nu$ at $\xi_\nu\simeq 80$, and the dynamo amplification $\cala_\kin\simeq 10^4$ is almost 3 orders of magnitude greater than the factor $\simeq 20$ due to compression.

After reaching $B_\nu$, the dynamo enters the nonlinear stage.
The nonlinear amplification factor is given by (eqs. \ref{eq:anl2} and \ref{eq:iq1})
\beqa
\cala_\non^2&\hspace{-0.3cm}=\hspace{-0.3cm}&1+2\chi\phiff\Gamma_{\nu0}\tffo\xi_\nu^{\frac 13-\frac12 (q_\epsilon+q_\nu)}I_{q_\epsilon-\frac 13}(\xi_\nu,\xi),
\label{eq:anl5}\\
&\hspace{-0.3cm}\simeq\hspace{-0.3cm} &1+\frac{0.088}{\frac 56-q_\epsilon}\left(\frac{\phiff \vtf^{3/2}}{h_f^{1/2} m_\sph'^{1/6}T_3^{1/2}n_0^{1/12}}\right)\xi_\nu^{q_\Gamma-\frac 12}\nonumber\\
& &~~~~~\times\left[1-\left(\frac{\xi_\nu}{\xi}\right)^{\frac 56-q_\epsilon}\right]~~~~~(\xi_\nu^{1/3}\gg 1),
\label{eq:anl6}
\eeqa
where we used equation (\ref{eq:qg}) for $q_\Gamma$ and equation (\ref{eq:xinu}) for $\xi_\nu$. This equation applies only for $\xi<\xi_{\max}$ since the dynamo cannot operate at higher densities.
In the absence of systematic variations in $T$ or $v_t$, 
we have $q_\epsilon=\frac 12$ and $q_\Gamma= \frac{5}{12}$, so $\cala_\non$ is typically $\sim 1$. For example, the case portrayed in Fig. 4 has $\cala_\non^2\simeq 1+0.75\phiff\xi_\nu^{-1/12}$ for $(\xi_\nu/\xi)^{1/3}\ll 1$.
As a result, 
the nonlinear amplification of
the field is primarily due to compression of the field.
The fact that $\cala_\non$ is smaller for simulations than for the physical case is expected since $\cala_\non\propto Re^{1/4}$ (see below equation \ref{eq:gtf}) and $Re$ is much smaller for simulations.

The dynamo reaches equipartition at $\xi_\eq$. However, just as in the case of $\xi_\nu$,
the uncertainty in $\cala_\kin$ means that we cannot predict the equipartition density or field in a simulation with any certainty. In equipartition, we have $\va=v_t$, so that
\beq
\xi_\eq=\left(\frac{v_t}{\vao\cala_\eq}\right)^6,
\label{eq:xieq2}
\eeq
where $\cala_\eq=\cala_\kin\cala_\non(t_\eq)$ is the amplification factor at the time that the field reaches equipartition. In Fig. \ref{fig:sim}a, we know all the parameters. For $\phiff=2$, for example, the field reaches  equipartition at $\nh\simeq 1.2\times 10^7$~cm\eee, when $B\simeq4\times 10^{-3}$~G. Keep in mind that these values are based on the assumption that $B_0=10^{-11}$~G; if the initial field were weaker, it would reach equipartition at a higher density with a correspondingly higher value of the field strength. The field then remains in equipartition and grows as $\nh^{1/2}$. As discussed in Section \ref{sec:gravity}, equipartition fields with $\phi_t\sim \frac 12$ result in mass-to-flux ratios $\mu_\Phi\sim 2-3$, which is small enough that magnetic fields can significantly affect star formation. Note that the full effect of this low mass-to-flux ratio is felt only in the central $3\times 10^{-3}$ of the core for $\xi_\eq\sim 10^{7}$ (equation \ref{eq:krho}), or about $100\,M_\odot$ for a minihalo with a gas mass of $4\times 10^4\,M_\odot$. If the field saturates at a value $\phi_\sat$ less than the equipartition value (equation \ref{eq:sat}), then it would saturate at a density $\phi_\sat^6$ less than that in equation (\ref{eq:xieq2}), corresponding to a mass $\phi_\sat^{-2.16}$ times greater; for $\phi_\sat=0.7$ \citep{federrath11a}, this is about a factor 2. 

We conclude that SPH simulations can follow a significant growth of the field in a gravitational collapse due to the action of a small-scale dynamo, but the mass in which the field reaches equipartition is small compared to the correct value and it is difficult to predict the final field in advance.
The results presented here will be compared with SPH simulations in Paper II.

\subsection{Grid-based Simulations of Mini-halo Dynamos}
\label{sec:predgrid}

Grid-based simulations of mini-halo dynamos are quite similar to SPH simulations, except that the numerical viscosity
is somewhat different (Appendix \ref{app:numerical}). Since the kinematic stage extends well into the gravitational
collapse due to the large value of the viscosity, the outer scale of the turbulence is the Jeans length, as for the SPH case. The Reynolds
number is then given by $Re=512/(64 J_{\max})^{4/3}$ (equation \ref{eq:reg3}), where $J_{\max}$ is the maximum value of the ratio of the grid size to the Jeans length allowed in the adaptive mesh simulation. The ratio of the magnetic Reynolds number to the critical value, $\rmcr=220/P_m^{1/2}$
(see the comment above equation \ref{eq:rmrmcr1}), is then
\beq
\frac{R_m}{\rmcr}=2.33P_m^{3/2}\left(\frac{1/64}{J_{\max}}\right)^{4/3}
\eeq
with the aid of Equation (\ref{eq:reg3}).
As discussed in Appendix \ref{app:numerical}, this implies that 
that the dynamo can operate ($R_m>\rmcr$) for $\lj/\Delta x>16-32$, as found
by \citet{federrath11b}, provided $P_m$ is in the range 1-2. 
More precisely, the dynamo can operate provided
\beq
J_{\max}<0.03P_m^{9/8},
\label{eq:jmax}
\eeq
which is 1/23 for our adopted value $P_m=1.4$. For a given grid size, 
$\Delta x$, the maximum density is the Truelove-Jeans density, $\rtj=\pi J_{\max}^2\cs^2/(G\Delta x^2)$ (equation \ref{eq:rtj}). Equation (\ref{eq:jmax}) then sets the maximum density for a dynamo to operate in a grid-based simulation, 
\beq
\nhm=1.23\times 10^{11}\left(\frac{P_m^{9/4}T_3}{\Delta x_{14}^2}\right)
~~~~\mbox{cm\eee},
\label{eq:nhm2}
\eeq
where $\Delta x_{14}=\Delta x/(10^{14}$~cm). The highest resolution in the grid-based simulation of Stacy et al. (in preparation) is $\Delta x=0.47\times 10^{14}$~cm. This gives $\nhm=1.2 \times 10^{12}$~ cm\eee, slightly less than the maximum density in their simulation. The fact that $\nhm$ is much larger in the grid-based simulation than in the SPH simulation of Stacy et al (in preparation) was by design: the grid-based simulation was a zoom-in on the cosmological SPH simulation. 

The dynamo amplification factor in the kinematic stage is given by equation (\ref{eq:akin}). 
Using the grid-based viscosity from equation (\ref{eq:nug2}), which has $\nu_g\propto v_t(T/\nh)^{1/2}$, we have $\Gamma_\nu\propto \calm \nh^{1/2}$ so that
\beq
\Gamma_{\nu}\tffo=6.93\left(\frac{1/64}{J_{\max}}\right)^{2/3}\calm\xi^{\frac 12},
\label{eq:gnu2}
\eeq
where $\calm=v_t/\cs$ is the turbulent Mach number.
In terms of $\avg{\calm}$, the weighted average value of the Mach number over the range of compression ratios from 1 to $\xi$, 
equations (\ref{eq:akin}) and (\ref{eq:iq00}) then imply
\beq
\cala_\kin=\exp\left[2.60\phiff \left(\frac{1/64}{J_{\max}}\right)^{2/3}\avg{\calm}I_{1/2}(1,\xi)\right],
\label{eq:akin4}
\eeq
where $I_{1/2}\simeq (2/3\pi)\ln\xi$ (equation \ref{eq:iq2}).
As in the case of SPH, the value of $\cala_\kin$ is very sensitive to the input parameters: Fig. \ref{fig:sim}b shows the significant differences resulting from a factor 2 difference in $\phiff$.
Just as in the case with SPH simulations,
grid-based simulations of gravitational collapse can
follow large amplifications of the field provided the resolution is high ($J_{\max}\la 1/64$), but the amplification
cannot be predicted in advance with any accuracy.
For the kinematic stage of the dynamo, an increase in resolution at a fixed value of $\xi$ is equivalent to an increase in $\phiff$ in determining the magnitude of the kinematic amplification: doubling the linear resolution (reducing $J_{\max}$ by a factor 2) is equivalent to increasing $\phiff$ by a factor $2^{2/3}$. The effects of an increase in resolution on a kinematic dynamo can thus be inferred from Fig. \ref{fig:sim}.

The logarithmic slope of $B(\rho)$ is given by equation (\ref{eq:dlnb3}). For grid-based simulations,
we have $q_\Gamma=\frac 12 + q_\calm$ from equation (\ref{eq:gnu2}), so that
\beq
\frac{d\ln B}{d\ln\rho}=\frac 23 + 0.55\phiff \calm_0
\left(\frac{1/64}{J_{\max}}\right)^{2/3}\xi^{q_\calm}.
\eeq
Note that the slope grows without bound as the resolution increases--i.e., as $J_{\max}$ and $\nu_g\propto J_{\max}^{4/3}$ decrease.
Indeed, as discussed in Section \ref{sec:predthy}, a viscosity as small as the actual viscosity allows the
kinematic dynamo to amplify the field by many orders of magnitude before the density changes significantly.

The dynamo leaves the kinematic stage of evolution when the field reaches the value
\beq
B_\nu=(4\pi\rho)^{\frac 12}v_\nu=1.11\times 10^{-7}\left(\frac{J_{\max}}{1/64}\right)^{\frac 13}\vtf\nh^{\frac 12}~~~~\mbox{G},  
\eeq
which is plotted in Fig. \ref{fig:sim}b. 
The discussion of the values of $\xi_\nu$ and $\xi_\eq$, which mark the onset of the nonlinear stage and reaching
equipartition, respectively, is similar to that in the previous section for the $\phiff=2,\, 4$ cases in SPH (for which $\nhm$ plays no role): The exponential uncertainty in $\cala_\kin$ implies that these quantities are essentially indeterminate in advance.
Of course, if one specifies the uncertain parameters, one can describe the kinematic dynamo accurately. For $\avg{\calm}=1$ and $\vtf=2$, one can show with the aid of equation (\ref{eq:akin4})  that $\cala_\kin(\xi_\nu)$ ranges from $10^{3.3}$ for $\phiff=1$ to $10^4$ for $\phiff=4$. The values of $\xi_\nu$ are $2.4\times 10^6$ and 100, respectively, so compression dominates dynamo amplification by an order of magnitude in the first case, but is relatively minor in the second.

We now consider the nonlinear evolution of the dynamo in a grid-based simulation.
From the discussion above equation (\ref{eq:gnu2}), we have $q_\Gamma=q_\calm+\frac 12$; simulations (e.g., \citealp{greifetal2012}, Stacy et al in preparation) show that the Mach number is approximately constant over a large range of densities in the collapse so that $q_\calm\sim 0$.
The nonlinear amplification factor
(equation \ref{eq:anl5}) then becomes
\beqa
\cala_\non^2&\hspace{-0.3cm}\simeq \hspace{-0.3cm}&1+\frac{0.18\phiff\calm_0}{\frac 56-q_\epsilon}
\left(\frac{1/64}{J_{\max}}\right)^{2/3}\xi_\nu^{q_\calm}\nonumber\\
&\hspace{-0.6cm}&~~~~~\times\left[1-\left(\frac{\xi_\nu}{\xi}\right)^{\frac 56-q_\epsilon}\right]~~~~~(\xi_\nu^{1/3}\gg 1).
\eeqa
The exponent $q_\epsilon\simeq \frac 12$ if there is no systematic variation
of velocity or temperature in the collapse ($q_v\simeq q_T\simeq 0$; see equation \ref{eq:qg}).
For grid-based codes, nonlinear dynamo amplification is small (as it is for SPH codes) provided the Mach number does not increase with
compression ($q_\calm\la 0$).
For the case shown in Fig. \ref{fig:sim}b ($\calm_0=1$, $q_\epsilon=\frac 12$, $q_\calm=0$, and $J_{\max}=1/64$), the amplification factor for the energy is $\cala_\non^2=1+0.55\phiff$. Equation (\ref{eq:xieq2}) then implies that the field reaches equipartition at $\xi_\eq\simeq (8\times 10^9,\,7\times 10^6,\, 5\times 10^4)$ for $\phiff=(1,2,4)$, respectively. If the field saturates at a value $\phi_\sat=0.7$ times smaller than the equipartition field \citep{federrath11a}, then these values are reduced by a factor 8.5. For a density power law
$\krho=2.2$, the field is saturated in the central $(11,140,800)M_\odot$, respectively. 

\subsubsection{Comparison with \citet{federrath11b}}

As noted above, uncertainties in the parameters prevent an accurate prediction of the amplification of the field in
the kinematic stage of the dynamo. However, once the simulation has been done, it is possible to compare our theoretical
estimates with the results of the simulation. Here we compare with the simulation of a kinematic dynamo in a gravitationally
collapsing cloud by \citet{federrath11b}. Their simulations covered the range $J_{\max}=1/8$ to $1/128$, and they found
dynamo action for $J_{\max}=1/32$ but not for 1/16. They presented their results in terms of the time normalized by the free-fall time,
$d\tau_F=dt/\tff$, so that (Eq \ref{eq:iq0})
\beq
\tau_F=\frac{1}{\tffo}\int_{t_0}^t \xi^{1/2}dt=\phiff I_{1/2}(1,\xi).
\eeq
(Note that their simulations did not include dark matter, so
$I_{1/2}\simeq(2/3\pi)\ln(64\xi)$ for $\xi^{1/3}\gg 1$.)
\citet{federrath11b} show that their results at late times imply $B/\rho^{2/3}\propto\cala_\kin$ varies as $\exp(\Omega\tau_F)$.
We find
\beq
\Omega=2.60\left(\frac{1/64}{J_{\max}}\right)^{2/3}\avg{\calm}
\label{eq:omega}
\eeq
from equation (\ref{eq:akin4}).
Over the normalized time interval from $\tau_F=8$ to $\tau_F=12$, 
the Mach number in the inner part of their simulation
increases by a factor 2 and has a typical value $\calm\simeq 0.5$. 
We therefore predict $\Omega\simeq 1.3[(1/64)/J_{\max}]^{2/3}$. 

How does this compare with their results? 
First of all, they find that $\cala_\kin\propto\exp(\Omega\tau_F)$ at late times, with $\Omega=\,\mbox{const}$
in a given simulation; we predict that $\Omega\propto\avg{\calm}$, which is nearly constant (their numerical
results imply $\avg{\calm}\simeq(\calm_0 \calm)^{0.5}\propto \xi^{0.05}$ approximately).
The values they found, $\Omega=0.4$ at $J_{\max}=1/64$ and 0.5 at $J_{\max}=1/128$,
are somewhat less than the values we predict. 
In agreement with their theoretical analysis, we predict that $\Omega \propto Re^{1/2}$ (eqs. \ref{eq:omega} and \ref{eq:reg3}), but as they point out, this does not agree with their numerical results, which are close to $\Omega\propto Re^{0.3}$ for constant $P_m$.
We note that our result follows from having the 
growth rate vary as $\nu^{-1/2}$ (Section \ref{sec:ssd}) and having the numerical viscosity for grid-based codes
vary as $\Delta x^{4/3}$ (Appendix \ref{app:numerical}), both of which appear reasonable. It is possible that the actual scaling of
$\Omega$ with $J_{\max}$ (or, equivalently, $Re$) appears only at higher resolution.

\section{Conclusions}

Magnetic fields affect the fragmentation of gravitationally collapsing gas, and that in turn affects the IMF. This is particularly important for the first stars since it determines  the nucleosynthesis that results when the stars explode as supernovae and whether Pop III stars can form with low enough masses that they survive today. As discussed in the Introduction, a great deal of work has been done on the origin of primordial magnetic fields. In the absence of any observational
data, their role in the formation of the first stars must come through theory and simulation. The aim of this paper has
been to make a theoretical estimate of the magnitude of the field in regions where the first stars formed and to
then compare that with the results that are expected from simulations, given that the numerical viscosity and
resistivity are orders of magnitude larger than the actual values. In a companion paper (Stacy et al in preparation), we describe the results
of a simulation of the formation of the first stars that includes magnetic fields.

As discussed in the Introduction, it has been realized for some time that small-scale dynamos can produce
dynamically important magnetic fields in regions of Pop III star formation. Dynamos require seed fields, and
a great deal of effort has gone into determining possible mechanisms for generating such fields. Mechanisms
that might have occurred in the early universe, such as those due to inflation or to phase transitions, are very
uncertain. The one mechanism that depends only on known physics is the Biermann battery \citep{bier50,bier51},
which can produce fields $\sim 10^{-24.5}$~G throughout the IGM after recombination \citep{naoz&narayan2013} and
$\sim 10^{-19}$~G in newly formed galaxies \citep{bier51}. Such fields must be amplified by small-scale dynamos
in a turbulent medium to become dynamically or observationally significant. Observations of $\gamma$-rays from blazars set
a lower limit of $10^{-17}$ G on intergalactic magnetic fields with a correlation length exceeding 1 Mpc, with larger values
for smaller correlation lengths \citep{nero10,tayl11}, although this result has recently been called into question \citep{brod18,alve19}.

The overall conclusion of our analysis is that a small-scale dynamo can amplify primordial fields created by the Biermann battery
mechanism to the point that the dynamo enters the nonlinear stage and that subsequent compression brings the field into approximate equipartition with the turbulent motions in the collapsing gas cloud. However, because the numerical viscosity is typically orders of magnitude greater that the actual value, the field in a simulation becomes dynamically significant in a much smaller mass than in reality.
We now separately summarize our results for the fields expected theoretically and those expected in numerical simulations.

(1) {\it The Biermann battery} generates weak magnetic fields ($\sim 10^{-4}\omega$, where $\omega=\curl\vecv$ is the vorticity) due to forces that produce unequal accelerations of the electrons and ions and have a curl, such as non-parallel pressure and density gradients. We confirmed the statement by \citet{kuls97} that dissipative processes in shocks do not significantly affect the operation of the Biermann battery. Standard estimates for the Biermann field are based on the vorticity produced by curved shocks on galactic scales and give values $\sim 10^{-19}$~G \citep{bier51,pudr89}, and we find a similar value for cosmic minihalos. We show that the subsequent turbulent cascade gives fields on the viscous scale in cosmic minihalos ($\sim 0.01$~pc) of order $10^{-16}$~G.

(2) {\it The small-scale dynamo.}  We summarized some of the key results on small-scale dynamos, which begin by amplifying fields on the viscous scale (or resisitive scale, if that is larger). Extensive theoretical work and simulations have shown that turbulence can amplify weak magnetic fields until they reach approximate equipartition (provided the magnetic Reynolds number, $R_m=Lv_L/\eta$, is large enough--see equation \ref{eq:rmcr}). For magnetic Prandtl numbers exceeding unity ($P_m=\nu/\eta>1$, where $\nu$ is the viscosity and $\eta$ is the resistivity), the largest fields are on subviscous scales until equipartition is reached on the viscous scale; we label that field $B_\nu$. Subsequently, both the magnitude and the scale of the field grow as it reaches equipartition with larger and larger eddies. In the post-recombination universe, ambipolar diffusion provides the dominant resistivity for fields $B\ga 10^{-13}\nh$~G (Appendix \ref{app:viscosity}). We followed the treatment of \citet{xu&lazarian2016} in treating non-ideal effects on the dynamo, summarizing their results on the complex behavior of the dynamo in two figures, one for the case of Ohmic resistivity (Fig. \ref{fig:ohm}) and one for resistivity due to ambipolar diffusion (Fig. \ref{fig:AD}). 
The field grows exponentially in the kinematic phase of the dynamo ($B<B_\nu$) and as $t^{1/2}$ in the nonlinear phase ($B>B_\nu$) provided $P_m(B_\nu)$ is not too small. The values of the parameters describing dynamos in minihalos are summarized in Table \ref{tab:dyn}.

(3) {\it Dynamos in a time-dependent medium.} The magnetic Reynolds number in a typical cosmic minihalo is large (see Table \ref{tab:dyn}), so flux freezing is a good approximation for the effects of compression. We determine the growth of the field
in a time-dependent medium due
both to compression, $B\propto\rho^{2/3}$, and to the dynamo. Because the growth rate of the field in the nonlinear stage of the dynamo is much less than that of the kinematic dynamo, compression generally dominates dynamo amplification of the field in the nonlinear stage. On the other hand, dynamo amplification is relatively more important for the specific magnetic energy, $\eb=B^2/8\pi\rho$, and as a result the nonlinear dynamo generally amplifies the magnetic field energy to the point that it is within an order of magnitude of equipartition in a gravitational collapse, even in the absence of compression.

(4) {\it Gravitational collapse.} In a CDM universe, the first stars form via the gravitational collapse of gas in a cosmic minihalo. 
In Appendix \ref{app:free} we first develop an approximation for the free-fall collapse of a constant density sphere;
our analytic expression for $r(t)$ is complementary to the approximation for $t(r)$ obtained by \citet{giri14}.
We then idealize the contraction of the baryons in the minihalo as a free-fall collapse of uniform density sphere of gas in a static dark-matter halo of constant density and show that the dark matter accelerates the collapse by slightly more than a factor 2.

(5) {\it Theoretically predicted magnetic field in the formation of the first stars.} 
The evolution of a dynamo in a collapsing minihalo depends on a large number of parameters: the initial density, $\nho$, the turbulent velocity, $v_t$ (which we parametrize in terms of the virial velocity,
$v_t=\phi_t v_\vir$), the temperature, $T$, the mass of the collapsing cloud, $M_0$, the rate of collapse (parametrized by $\phiff$), and the rate at which these quantities vary with density (denoted by $q_x$ for quantity $x$). (The initial value of the field, $B_0$, enters only logarithmically, and is important only if it is many orders of magnitude less than our estimate of $\sim 10^{-16}$~G.) Choosing values of these parameters that are consistent with simulations (e.g., those of \citealp{greifetal2012}), we
find that the time for the field to grow from its initial amplitude
$\sim 10^{-16}$~G to equipartition at the viscous scale, $B_\nu\sim 10^{-8}$~G, is less than the virial time in the minihalo; hence,
the exponential growth of the field occurs at approximately constant gas density. This rapid growth of the field is consistent with that found in previous work (e.g., \citealp{schleicheretal2010,schoberetal2012b}).
The subsequent nonlinear 
dynamo amplification is sufficient to bring the field energy to within about an order of magnitude of equipartition; nonetheless, the overall amplification of the field is generally dominated by compression. 
We estimate that the field first reaches equipartition with turbulent velocities of order 2 km s\e\ (taken from simulations)
at a value $\sim 10^{-4}$~G; the field subsequently grows as $\nh^{1/2}$.
The field reaches equipartition with the central 5\% of the mass of the gas. Our conclusion that the field reaches equipartition in a minihalo at
$z\sim 25$ differs from that of \citet{xu&lazarian2016}, who found that equipartition was not reached until a time of about $6\times 10^8$ yr (the age of the universe at $z\simeq 8$) since they did not consider the increase in density that occurs in star formation. 

(6) {\it Magnetic effects on the first stars.} The ratio of
the mass-to-flux ratio to the critical value, $\mu_\Phi$, is predicted to be about 2-3. Magnetic fields in contemporary star formation regions are also in approximate equipartition and have similar values of $\mu_\Phi$ \citep{crutcher12}, so magnetic fields could play an important role in the formation of the first stars. The fields in regions of first-star formation were produced in a turbulent small-scale dynamo and lack large scale order, in contrast to those in regions of contemporary star formation, and as a result protostellar outflows are unlikely from the first stars.

We then discussed the possible outcome of simulations of the growth of magnetic fields in the formation of a primordial star in
a minihalo, using either an SPH or a grid-based ideal MHD code. The viscosity and resistivity in the simulations are assumed to be purely numerical.

(1) {\it Numerical viscosity and resistivity.} We developed a method of estimating the numerical viscosity, $\nu$, that is in agreement with the estimate of \citet{benz08} for grid-based codes and of \citet{baue12} for SPH codes. The value of the numerical viscosity in current simulations is typically more than 1000 times greater than the actual viscosity in weakly ionized primordial gas. We estimate that the magnetic Prandtl number is $P_m=\nu/\eta\sim 1.4$ for grid-based codes based on the results of \citet{federrath11b}; we adopt the same value for SPH codes.

(2) {\it Suppression of the dynamo by numerical resistivity.} Dynamos cannot operate if the magnetic Reynolds number, $R_m$, is too small. We determined the maximum density, $\nhm$, at which dynamos can operate for both SPH and grid-based AMR codes under the assumption that the length scale in the Reynolds number is set by the Jeans length (eqs. \ref{eq:nhmax} and \ref{eq:nhm2}). Low values of $\nhm$ lead to high values of the density at which the field reaches equipartition and therefore small fractions of the collapsing mass in which the field is dynamically significant. 

(3) {\it Predicted magnetic fields in simulations of gravitational collapsing gas.} The large value of the numerical viscosity for a resolution of 64 cells per Jeans length ($J=1/64$), a typical value in current simulations, implies that the growth rate of the kinematic dynamo is $\la 1/30$ of the physically correct value. As a result the growth of the field by compression is predicted to exceed than that due to the dynamo if the collapse occurs at the free-fall rate ($\phiff\simeq 1$). After the dynamo enters the nonlinear stage, dynamo amplification is predicted to be relatively less important compared to compression in simulations than in reality. As noted above, the evolution of the dynamo depends on a number of parameters; in simulations, the resolution is an additional important parameter. The total amplification in the kinematic stage of the dynamo is exponentially dependent on these parameters, so the growth of the field in a simulation is difficult to predict in advance. Examples of the predicted outcomes of simulations of the growth of magnetic fields in a gravitationally collapsing cloud are given in Fig. \ref{fig:sim}. Increasing the resolution of the simulation increases the mass fraction in which the field can reach equipartition.

\section*{Acknowledgements}
We thank Siyao Xu for extensive discussions on her work and for comments on drafts of this paper. 
We thank Robi Banerjee, Eric Blackman, Christoph Federrath, Robert Fisher, Alex Lazarian, Alex Schekochihin, Zack Slepian and  Volker Springel for helpful comments and Andrew Cunningham for sharing data analysis routines with us. We also thank the referee, whose recommendations significantly improved the paper.
This research was supported in part by the NSF though grant AST-1211729 and by NASA through ATP grants NNX13AB84G and NNX17AK39G.

\section*{Data Availability Statement}

No new data were generated or analysed in support of this research.

\bibliographystyle{mnras}
\bibliography{bfield}{}

\appendix

\section{Viscosity and Resistivity of Primordial Gas}
\label{app:viscosity}

\subsection{Viscosity}

For a primordial gas with $n_{\rm He}/\nh\la 0.1$, the viscosity is very close to that of atomic hydrogen.
A fit to the results of \citet{vran13} for the dynamic viscosity of atomic hydrogen
based on the measured cross section for H-H scattering gives:
\beq
\eta_{\rm visc,\,HH}=1.14\times 10^{-5}T_3^{0.84}~~~\mbox{kg s\e\ m\e}.
\label{eq:visc}
\eeq
They quote (3.95, 5.5, 8.6)$\times 10^{-5}$ kg s\e\ m\e\ at $T=(4400,\,6560,\, 11150)$ K,
whereas the fit gives (3.96,\,5.53,\,8.64), for excellent agreement. There are no data for $T=1000$~K, but the cross section for H-H scattering, $\sigma_{\rm HH}$,
continues to rise and $\eta_{\rm visc,\,HH}\propto 1/\sigma_{\rm HH}$ continues to fall as the energy decreases,
consistent with the behavior of equation (\ref{eq:visc}). In cgs units, the viscosity is 10 times larger. We adopt a helium abundance $n_{\rm He}/\nh=1/12$, a good approximation to the most recent value ($1/12.20$) from Big Bang nucleosynthesis combined with observations of the cosmic microwave background \citep{fiel19}. The mass per H nucleus is then $\muh=2.23\times 10^{-24}$ g, and
the kinematic viscosity is 
\beq
\nu\simeq\frac{\eta_{\rm visc,\,HH}}{\rho}=\frac{\eta_{\rm visc,\,HH}}{\nh\muh}=5.11\times 10^{19}\,\frac{T_3^{0.84}}{\nh}~~~\mbox{cm$^2$ s\e}.
\label{eq:nu}
\eeq

\subsection{Ambipolar Resistivity}

The ambipolar resistivity (in the terminology of \citealp{pint08a}) is
\beq
\ead=\frac{B^2}{4\pi\rho_i\nu_{in}}=\frac{B^2}{4\pi\rho_i\rho_n\gad},
\label{eq:ead1}
\eeq
where $\rho_i$ is the mass density of ions, $\nu_{in}$ is the ion-neutral collision frequency,
and the collisional drag coefficient, $\gad$, is defined through
\beq
\rho_i\nu_{in}=\rho_i\rho_n\gad.
\eeq
The expression for $\ead$ follows from balancing the drag force, $\rho_i\nu_{in}v_d$, where $v_d$ is the relative ion-neutral velocity,
with the Lorentz force, $ B^2/4\pi\ell_B$, where $\ell_B$ is the length scale over which the field varies, and then
setting $\ead\sim \ell_B v_d$ (for an actual derivation, see \citealp{bran94} or \citealp{pint08a}). In our case, there is one dominant
ion, H$^+$, and (prior to molecule formation) two dominant neutrals, H and He. For low ionization, H and He will
have the same velocity, so that
\beq
\rho_i\nu_{in}=\Sigma_j n_i n_j\mu_{ij}\avg{\sigma v}_{ij}
\eeq
\citep{glas05}, where the sum is over the neutrals and $\mu_{ij}$ is the reduced mass. Because the H$^+$-He collision rate is 
only about 1/6 of the H$^+$-H collision
rate \citep{pint08b} and the He abundance is low ($x_\he\equiv n_{\rm He}/\nh=1/12$), H$^+$-He collisions
make a negligible contribution to the ion-neutral collision rate. Under the assumption that the ionization is very small, 
we then have
\beq
\nu_{in}=\frac 12 \nh\avg{\sigma v}_\hhp.
\eeq
The neutral density is $\rho_n=(1+4x_\he)\nh m_{\rm H}$, so that
\beq
\gad=\frac{\nu_{in}}{\rho_n}=\frac{\avg{\sigma v}_\hhp}{2(1+4x_\he)m_{\rm H}}.
\eeq
In the text, we also need the neutral-ion collision frequency, $\nuni$, which satisfies $\rho_n\nuni=\rho_i\nu_{in}$, so that
$\nuni=\rho_i\gad.$
\citet{glas05} modified \citet{drai80}'s determination of the rate coefficient for H-H$^+$ collisions, obtaining
$\avg{\sigma v}_\hhp=2.13\times 10^{-9}v_{\rm rms,5}^{0.75}$ cm$^3$ s\e\  for $v_{\rm  rms}> 1$~km~s\e,
which leads to
\beq
\gad= 6.36\times 10^{14}\left(\frac{v_{\rm rms,5}^{0.75}}{1+4x_\he}\right)~~~\mbox{cm$^3$ s\e\ g\e}.
\eeq
For $x_\he=0.1$, this agrees with the result of \citet{glas05}; for $x_\he=1/12$, this gives $\gad=4.77\times 10^{14}v_{\rm rms,5}^{0.75}$
cm$^3$ s\e\ g\e.

To express $\gad$ in terms of the temperature, we note that
for two species, $s$ and $s'$, with Maxwellian velocity distributions moving at a relative velocity $v_d$,  we have
\beq
v_{\rm rms}=\left(v_d^2+\frac{8kT_{ss'}}{\pi\mu_{ss'}}\right)^{1/2},
\label{eq:vrms}
\eeq
where 
\beqa
T_{ss'}&\hspace{-0.3cm}=\hspace{-0.3cm}&\frac{m_{s'}T_s+m_sT_{s'}}{m_s+m_{s'}}\rightarrow T,\\ \mu_{ss'}&\hspace{-0.3cm}=\hspace{-0.3cm}&\frac{m_sm_{s'}}{m_s+m_{s'}}\rightarrow
\frac 12 m_{\rm H},
\eeqa
(e.g., \citealp{pint08b}) and where the simplified results apply to an H-H$^+$ plasma.\footnote{Note that for $v_d=0$, $v_{\rm rms}$ is actually the mean particle velocity, not the rms velocity, but we follow the notation of \citet{pint08b} here.}
Expressing $v_{\rm rms}$ as
\beq
v_{\rm rms}=\left(\frac{8kT_{ss'}}{\pi\mu_{ss'}}\right)^{1/2}\phi_d^{4/3},
\eeq
we have for H-H$^+$ collisions
\beqa
v_{\rm rms}&\hspace{-0.3cm}=\hspace{-0.3cm}&6.48\times 10^5 \,\phi_d^{4/3}T_3^{1/2}~~~\mbox{cm s\e},\\
\phi_d&\hspace{-0.3cm}=\hspace{-0.3cm}&\left[1+\left(\frac{v_{d,5}}{6.48}\right)^2\frac{1}{T_3}\right]^{0.375},
\label{eq:phid}\\
\hspace{-1cm}\avg{\sigma v}_\hhp&\hspace{-0.3cm}=\hspace{-0.3cm}&8.65\times 10^{-9}\phi_d T_3^{0.375}~~~\mbox{cm$^3$ s\e},\\
\gad&\hspace{-0.3cm}=\hspace{-0.3cm}&1.94\times 10^{15}\,\phi_d T_3^{0.375}~~~\mbox{cm$^3$ s\e\ g\e},\\
\nuni&\hspace{-0.3cm}=\hspace{-0.3cm}&\rho_i\gad,\\
&\hspace{-0.3cm}=\hspace{-0.3cm}&3.24\times 10^{-13} \,\phi_d \xif\nh T_3^{0.375}~~~\mbox{s\e},
\label{eq:nuni}
\eeqa
where $\phi_d$ is determined from equation (\ref{eq:vrms}), the final two expressions  are for $x_\he=1/12$,
and $\xif=(n_i/\nh)/10^{-4}$ is the normalized ionization fraction.
%
Our result for $\gad$ is larger than that of \citet{xu&lazarian2016} since we used the value of
$\avg{\sigma v}$ given by \citet{glas05} instead of that by \citet{drai83}; in addition, the value adopted by \citet{xu&lazarian2016}
appears to be for the case of molecular clouds, for which the dominant ions are heavy molecules such as HCO$^+$.

Since the magnetic field and therefore the ambipolar resistivity, $\ead\propto B^2$,
vary by orders of magnitude, it is convenient to express $\ead$ in normalized form. Normalizing the \alfven\ velocity with respect
to the turbulent velocity on large scales, $v_t$, and the field relative to the equipartition value at the viscous scale, $B_\nu$, 
(equation \ref{eq:bnu}), we have
\beqa
\hspace{-0.6cm}\ead&\hspace{-0.3cm}=\hspace{-0.3cm}&3.08\times 10^{22}\left(\frac{\vtf^2}{\phi_d\xif \nh T_3^{0.375}}\right) \frac{\va^2}{v_t^2}~~\mbox{cm$^2$ s\e},\\
&\hspace{-0.3cm}=\hspace{-0.3cm}& 3.96\times 10^{20}\left(\frac{T_3^{0.04}\vtf^{3/2}}{\phi_d\xif \nh^{3/2}r_\pc^{1/2}}\right)\frac{B^2}{B_\nu^2}~~\mbox{cm$^2$\,s\e}.
\eeqa
Alternatively, in terms of $\beta=8\pi\rho\cs^2/B^2=2\cs^2/\va^2$, we have
\beq
\ead=\frac{2\cs^2}{\beta\rho_i\gad}=4.13\times 10^{23}\left(\frac{T_3^{0.62}}{\phi_d\xif\nh\beta}\right)~~~\mbox{cm$^2$ s\e}.
\eeq

\subsection{Ohmic Resistivity}
\label{app:ohm}

As noted by previous authors (e.g., \citealp{kuls92}), the Ohmic resistivity is generally negligible compared to the AD resistivity unless the field is very weak:
Since the drag due to ion-neutral collisions is much greater than that due to electron-neutral collisions, the Ohmic resistivity is determined by electron-ion and electron-neutral interactions \citep{pint08a},
\beq
\eta_{\rm O}=\frac{c^2}{4\pi}\left(\frac{m_e}{e^2n_e}\right)\left(\nu_{ei}+\nu_{en}\right),
\eeq
where 
\beq
\nu_{ss'}=\left(\frac{m_{s'}}{m_s+m_{s'}}\right) n_{s'}\avg{\sigma v}_{ss'}
\eeq
is the collision rate for momentum transfer between particles of type $s$ and those of type $s'$.
(We follow \citealp{pint08a} in writing $\eta_{\rm O}=c^2/(4\pi\sigma_{\rm cond})$ for the Ohmic resistivity, 
where $\sigma_{\rm cond}$ is the electrical conductivity.)
\citet{pint08b} give
\beqa
\hspace{-0.7cm}\avg{\sigma v}_{e{\rm H}^+}\hspace{-0.3cm}&=&\hspace{-0.3cm}\frac{2.30\times 10^{-3}}{T_3^{3/2}}\left(\frac{\ln\Lambda}{20}\right)~\mbox{cm$^3$ s\e},\\
\avg{\sigma v}_{e{\rm H}}\hspace{-0.3cm}&=&\hspace{-0.3cm} 1.41\times 10^{-7}T_3^{0.6}\exp\left(-0.43 T_3^{1/2}\right)~\mbox{cm$^3$ s\e},~~~~~
\eeqa
where $\ln\Lambda$ is the Coulomb logarithm and where
we have assumed that the drift velocity of the electrons relative to the neutrals is much less than 100 km s\e\ in
the second expression. As a result, we have
\beqa
\eta_{\rm O}=\left[\frac{6.5\times 10^8}{T_3^{3/2}}\left(\frac{\ln\Lambda}{20}\right)\right.+~~~~~~~~~~~~~~~~~~~~~~~~~~~~~~~~\\
~~~~~~~~~\left.\frac{4.0\times 10^8 T_3^{0.6}}{\xif}\,\exp\left(-0.43 T_3^{1/2}\right)\right]~~~\mbox{cm$^2$ s\e},
\eeqa
which is negligible compared to $\ead$ for $\nh\beta\ll 10^{15}$ cm\eee. 
More specifically, ambipolar diffusion dominates electron-ion Ohmic resistivity and electron-neutral Ohmic resistivity for
\beqa
\hspace{-0.7cm}B\hspace{-0.3cm}&>&\hspace{-0.3cm}8.9\times 10^{-14}\left[\frac{\xif\phi_d}{T_3^{1.12}}\left(\frac{\ln\Lambda}{20}\right)\right]^{1/2}\nh~~\mbox{G},\\
\hspace{-0.3cm}&>&\hspace{-0.3cm}7.0\times 10^{-14}\phi_d^{1/2}T_3^{0.49}\exp\left(-0.22T_3^{1/2}\right)\nh~~\mbox{G},~~~~~
\eeqa
respectively. Hence, ambipolar diffusion is typically dominant for $B\ga 10^{-13}\nh$~G.

\section{Free-fall Collapse}
\label{app:free}

Gravitational collapse is often described approximately by the collapse of a uniform, pressureless sphere of gas, which
has the parametric solution \citep{spit68}
\beqa
r&=&r_0\cos^2\psi,\\
\psi+\frac 12 \sin\ 2\psi&=&\frac{\pi}{2}\left(\frac{t}{\tffo}\right),
\eeqa
where $\tffo=(3\pi/32G\rho_0)^{1/2}=1.41\times 10^{15}n_{\rm H,0}^{-1/2}$~s is the initial free-fall time of the gas--i.e., the time at which a cloud beginning at rest with a radius $r_0$ collapses to a singularity. In cosmology, this is the tophat solution.
\citet{giri14} have shown that it is possible to obtain an accurate approximation
for the time as a function of the radius for free-fall collapse; unfortunately, solving this relation for the radius as a function
of time does not give an accurate result at late times. Instead, one can show that in a free-fall collapse,
gas that is initially static at a radius $r_0$ is at a radius 
\beq
r=\phi_r r_0(1-\tau^2)^{2/3}
\label{eq:rapp}
\eeq
at a time $t$,
where $\tau\equiv t/\tffo$. The factor $\phi_r\rightarrow 1$ for
$\tau\rightarrow 0$ and $\phi_r\rightarrow (3\pi/8)^{2/3}=1.115$ for $\tau\rightarrow 1$. The approximation
$\phi_r\simeq 1.05$ is accurate to within 6\% for all $\tau$ between 0 and 1. An  approximation that is accurate to within 0.3\% for all $\tau$ in this range is
\beq
\phi_r\simeq \left[0.234+0.766\left(1-\tau^{3/2}\right)^{2/3}\right]^{-0.075}.
\eeq
The normalized density is 
\beq
\xi\equiv \frac{\rho}{\rho_0}=\left(\frac{r_0}{r}\right)^3=\frac{1}{\phi_r^3(1-\tau^2)^2}.
\label{eq:rho}
\eeq
Taking $\phi_r=(1,1.05)$ gives an accuracy of (40\%,\,20\%) for the density, respectively; taking $1/\phi_r^3=(8/3\pi)^2=0.72$ is accurate to 10\% for $\rho>100\rho_0$.
The time is given by
\beq
\tau=\left[1-\left(\frac{\rho_0}{\phi_r^3\rho}\right)^{1/2}\right]^{1/2}\rightarrow 1-\frac{4}{3\pi}\left(\frac{\rho_0}{\rho}\right)^{1/2},
\label{eq:tau}
\eeq
where the final step gives an accuracy for $1-\tau$ that is better than 10\% for $\rho> 100\rho_0$. 

In mini-halos, dark matter is initially dominant, so
we generalize the treatment above to allow for this. In addition, we allow for the possibility that the collapse occurs at a rate $\phiff$ less than free fall due to the fact that real collapses are not pressureless.
The equation of motion for
a shell of gas at radius $r$ inside a collapsing cloud is then
\beq
\frac{dv}{dt}= - \frac{1}{\phiff^2}\left[\frac{GM(r)}{r^2}+\frac{4\pi G\rho_d r}{3}\right],
\eeq
where 
$M(r)$ is the mass of gas inside $r$ and the numerical factor $\phiff\geq 1$
in the absence of external compression, since the gas pressure resists collapse.
We assume that the density of dark matter, $\rho_d$, is spatially constant and remains constant in time; that is, we neglect the adiabatic compression of the dark matter, and we assume that the free-fall time is much less than the age of the universe. Note that inside the cloud, we have $M(r)
\propto r^3$, so that $dv/dt\propto r$ and the collapse 
of a constant-density sphere in a constant-density background is homologous, just as in the case with no dark matter. The solution of this equation is
\beq
v^2=v_g^2\left[\frac{1}{y}-1+f_{db}(1-y^2)\right],
\label{eq:v2}
\eeq
where $y\equiv r/r_0$, 
\beq
v_g\equiv\frac{1}{\phiff}\left[\frac{2GM(r_0)}{r_0}\right]^{1/2},
\label{eq:vg}
\eeq 
and
\beq
f_{db}\equiv \frac{\Omega_d}{2\Omega_b},
\eeq
which is $f_{db}=3.25$ for the parameters adopted in the text. 
At late times in the collapse, when $y\ll 1$, we have
\beq
v^2\simeq \frac{2GM(r)}{\phiff^2r},
\eeq
since $M(r)=M(r_0)$. This relation can be used to determine the value of $\phiff$ in a simulation.

Since the time for the gas to collapse to infinite density in the absence of dark matter is now $\phiff\tffo$, we generalize the definition of $\tau$ to
\beq
\tau=\frac{t}{\phiff\tffo}.
\eeq
Note that $\tffo=(3\pi/32 G\rho_0)^{1/2}$ is the initial free fall time for the gas alone.

In the text we need the integral of $\xi^q$ over time,
\beqa
I_q(\xi_1,\xi_2)\hspace{-0.3cm}&=&\hspace{-0.3cm}\frac{1}{\phiff\tffo}\int_{t_1}^{t_2}\xi^qdt=\int_{\tau_1}^{\tau_2}\xi^q d\tau,
\label{eq:iq0}\\
\hspace{-0.3cm}&=&\hspace{-0.3cm}\frac{2}{3\pi}\hspace{-0.1cm}\int_{\xi_1}^{\xi_2}\hspace{-0.2cm}\frac{\xi^{q-\frac 32}d\xi}{(1-\frac{1}{\xi^{1/3}})^{1/2}[1+\frac{f_{db}}{\xi^{1/3}}(1+ \frac{1}{\xi^{1/3}})]^{1/2}},~~~~~~
\label{eq:iq}
\eeqa
where we used $dt=r_0dy/v$, $y=\xi^{-1/3}$ and $r_0/v_g\phiff\tffo=2/\pi$. 
This expression is exact; it is not based on the approximate result for $r(t)$ given above.
For $q=0$, this gives 
\beq
t(\xi)=\phiff\tffo I_0(1,\xi)
\eeq
and therefore $t(r)$ since $r=r_0\xi^{-1/3}$.
Note that the effect of dark matter, which is parametrized by the factor $f_{db}$,
becomes negligible at small radii (large $\xi$). Note also that for large $\xi$, $I_q$ is proportional to $\xi^{q-1/2}$: 
the range of time integration scales as
the free-fall time, $\tff\propto\xi^{-1/2}$.

It is now possible to determine the collapse time of the gas in the presence of static dark matter, $\tcoll$.
For $q<\frac 12$, define
$I_{q,\infty}=I_q(1,\infty)$.
For $q=0$, numerical evaluation of the integral in equation (\ref{eq:iq})
gives the collapse time based on the total amount of matter, $\tcoll$:
\beq
I_{0,\infty}=\int_0^\tcoll \frac{dt}{\phiff\tffo}=\frac{\tcoll}{\phiff\tffo}=0.46
\eeq
for $f_{db}=3.25$. 
In the absence of dark matter, one can show that $I_{0,\infty}=1$ as it should: for $f_{db}=0$, the collapse time is $\tcoll=\phiff\tffo$, as noted above.

We now consider the particular case in which the integration extends from the initial density
($\xi_1=1$) to a large density ($\xi_2\gg 1$) for $f_{db}=3.25$.
For $q<\frac 12$, we have
\beq
I_q(1,\xi_2)\simeq I_{q,\infty}-\frac{2}{3\pi(\frac 12-q)}\,\xi_2^{-\left(\frac 12 -q\right)}~~~~~(\xi_2^{1/3}\gg 1),
\label{eq:iq1}
\eeq
where $I_{q,\infty}$ must be evaluated numerically.
For example, for $q=-\frac 12$, $I_{q,\infty}=0.278$; for $q=\frac 16$, $I_{q,\infty}=0.639$; and 
for $q=\frac{5}{12}$, $I_{q,\infty}= 2.43$. 
The approximation
\beq
I_q(1,\infty)=I_{q,\infty}\simeq\frac{0.47}{(1-2q)^{0.87}}
\label{eq:iqapp}
\eeq
is accurate to within 10\% for the range $-\frac 12<q<\frac{5}{12}$.
For $q\geq \frac 12$, $I_q(1,\xi)$ diverges at large $\xi$.

For $q=\frac 12$, an approximation for $I_{1/2}(1,\xi_2)$ that 
is accurate to within about 1\% is
\beq
\frac{2}{\pi}\ln\left\{1+\frac {2\left(\xi_2^{\frac 13}-1\right)^{\frac 12}}{1+f_{db}'}\left[\left(\xi_2^{\frac 13}+f_{db}'\right)^{\frac 12}+\left(\xi_2^{\frac 13}-1\right)^{\frac 12}\right]\right\}
\label{eq:iq2}
\eeq 
with $f_{db}'=f_{db}[2(1+\xi_2^{-1/3})]^{1/2}$.
For $\xi_2^{1/3}\gg 1$, $I_{1/2}\rightarrow (2/3\pi)\ln\xi_2$.

For $q>\frac 12$, we have
\beq
I_q(1,\xi_2)\simeq \frac{2}{3\pi(q-\frac 12)}\left(\xi_2^{q-\frac 12}-1\right)~~~~~(\xi_2^{1/3}\gg 1).
\label{eq:iq3}
\eeq

Finally, in order to treat small-scale dynamos in collapsing gas clouds with no dark matter, one needs to know the values of $I_q$ in this case as well. For $q=\frac 12$, the value of $I_q$ is given by equation (\ref{eq:iq2}) with $f_{db}'=0$; for $q>\frac 12$, equation (\ref{eq:iq3}) applies as is. For $q<\frac 12$, equation (\ref{eq:iq1}) applies with
\beq
I_{q,\infty}=\frac{2}{3\surd\pi(\frac 12-q)}\,\frac{\Gamma(\frac 52-3q)}{\Gamma(2-3q)}.
\eeq

\section{Numerical Viscosity and Resistivity}
\label{app:numerical}

Here we estimate the numerical viscosity in both grid-based and SPH codes. We begin by presenting a
method of determining the numerical viscosity for subsonic turbulence based on
the fact that viscosity
suppresses the $k^{-5/3}$ energy spectrum of Kolmogorov turbulence by a factor  \citep{pope00}
\beq
f(k\ell_\nu)\simeq \exp\left(-5.2\left\{[(k\ell_\nu)^4+0.4^4]^{1/4}-0.4\right\}\right),
\label{eq:pope}
\eeq
where $k$ is the wavenumber, $\ell_\nu=(\nu^3/\epsilon)^{1/4}$ is the viscous scale (equation \ref{eq:ellnu}), and $\epsilon=v_\ell^3/\ell$ is
the constant energy flux in the turbulence. \citet{pope00} showed that this is in good agreement with experimental
data and \citet{baue12} have shown that it accurately describes the turbulent
energy spectrum calculated with the {\sc arepo} code (with the exception of the bottleneck effect, which is absent from the result of
\citealp{pope00}), in both its fixed grid and moving mesh versions.
Numerical viscosity is not exactly equivalent to a physical viscosity. One manifestation of this is that turbulence simulations without a physical viscosity
show a larger bottleneck effect than those that solve the Navier-Stokes equations and resolve the dissipation 
range (V. Springel 2019, private communication). Another is that the effective Reynolds number in simulations of turbulent mixing is problem dependent
\citep{leco16}. Nonetheless, as shown by the excellent agreement \citet{baue12} found between their turbulence simulations and 
equation (\ref{eq:pope}), that equation provides a reasonable basis for estimating the effective numerical viscosity.

Equation (\ref{eq:pope}) predicts that viscosity has a substantial effect on the turbulence when
$f=\frac 12$, which occurs at $\kh\ell_\nu=0.485\simeq 0.5$. Since the viscosity is 
$\nu=\ell_\nu^{4/3}\epsilon^{1/3}$, it follows that
\beq
\nu=0.40\left(\frac{\epsilon}{\kh^{4}}\right)^{1/3}
=0.034\left(\frac{\epsilon L^4}{{\kh'}^4}\right)^{1/3}.
\label{eq:nukh}
\eeq
where we have also expressed the viscosity in terms of
the normalized wavenumber, $k'=kL/2\pi$, which ranges from 1 to $\caln_g$ in grid-based simulations and which is often used in reporting the results of simulations.
Since \citet{pope00}'s expression does not include the bottleneck effect, that effect must be eliminated in evaluating $\kh$.

We validate this approach by comparing with the results of \citet{baue12}. They carried out a simulation with the {\sc arepo} code
with a sound speed $\cs=1$,
Mach number $\calm=0.3$, and a box size $L=1$, so that $\epsilon=v_L^3/L=0.3^3=0.027$. The simulation corresponded to 
 $\caln_g=256$ cells in each direction, and the total (physical plus numerical) viscosity was $\nu=1.5\times 10^{-4}$.
 After removing the bottleneck effect apparent in their results, we estimate $\kh\simeq 140$ from their plot of the velocity
 power spectrum--i.e., 
 the normalized power spectrum at $k=140$ is half the value it has at $k=2\pi$. 
 (In terms of $k'$, their results show that the normalized power spectrum at $k'=22$ is half the value it has at $k'=1$).
 According to
 equation (\ref{eq:nukh}), this corresponds to a viscosity $\nu=1.65\times 10^{-4}$, in excellent agreement with
 their value in view of the uncertainty in the estimate of $\kh$.

\subsection{Grid-based Codes}
\label{app:gridcodes}

First consider grid-based codes, which have cells of size $\Delta x=L/\caln_g$.
The numerical viscosity in the grid-based {\sc flash} code has been evaluated by \citet{benz08}
through analysis of the longitudinal structure function, and
was found to correspond to $\ell_\nu\simeq 0.6\Delta x$. It follows that the numerical viscosity for grid-based codes is
\beqa
\nu_g&=&\ell_\nu^{4/3}\epsilon^{1/3}=\frac{\ell_\nu^{4/3}v_L}{L^{1/3}}\simeq 0.5 v_L\Delta x\left(\frac{\Delta x}{L}\right)^{1/3}\hspace{-0.2 cm},
\label{eq:nugrid1}\\
&=& 0.5\left(\frac{v_L\Delta x}{\caln_g^{1/3}}\right),
\label{eq:nugrid2}
\eeqa
where $v_L$ is the velocity on the scale $L$. More precisely, $v_L=(\epsilon L)^{1/3}$, where $\epsilon$ is the specific
energy dissipation rate (equation \ref{eq:vell}); while it is comparable to the rms turbulent velocity, $v_t$, in a simulation, there is no assurance that
the two velocities are equal. Nonetheless, since $v_t$ is generally the only global velocity quoted in
simulations, we shall use it in estimating the numerical viscosity.

Although equation (\ref{eq:pope}) was obtained for incompressible hydrodynamic turbulence, it works for supersonic
turbulence and MHD turbulence as well. (However, the results for the viscosity are valid only for subsonic turbulence
since they are based on Kolmogorov scaling.) Noting that $\Delta x =L/\caln_g$, we have
\beq
\frac{\ell_\nu}{\Delta x}=\frac{\kh \ell_\nu}{\kh\Delta x} \simeq \frac{0.5}{2\pi\kh' \Delta x/L}=
\frac{\caln_g}{4\pi\kh'}
\label{eq:ellx}
\eeq
for $\kh\simeq 0.5/\ell_\nu$. 
We estimate $\kh'=150$ for the Mach 5.5 simulation on a $1024^3$ grid by \citet{fede10}, which gives
$\ell_\nu=0.54\Delta x$. 
For the MHD simulation with a sonic Mach number of 10 and an \alfven\ Mach number of $\surd 5$ on a $512^3$ grid by \citet{lietal2012}, we
estimate $\kh'=62$, corresponding to $\ell_\nu=0.66\Delta x$. In both cases, these results are quite close to the value found by \citet{benz08}.
The corresponding result for {\sc arepo} is
$\ell_\nu=0.9\Delta x$, which is larger than the other values because it included a physical viscosity.
For the value we adopt, $\ell_\nu=0.6\Delta x$ \citep{benz08}, we have $\kh'=\caln_g/(2.4\pi)=\caln_g/7.5$.

The Reynolds number based on equation (\ref{eq:nugrid2}) is
\beq
Re=\frac{Lv_L}{\nu_g}=\frac{Lv_L}{0.5 v_L\Delta x/\caln_g^{1/3}}=2\caln_g^{4/3}.
\label{eq:renum}
\eeq
This result can also be derived directly from equation (\ref{eq:re}):
\beq
Re=\left(\frac{L}{\ell_\nu}\right)^{4/3}=\left(\frac{\caln_g}{\ell_\nu/\Delta x}\right)^{4/3},
\label{eq:renum2}
\eeq
which is $1.98\caln_g^{4/3}$ for $\ell_\nu=0.6\Delta x$. By contrast, \citet{federrath11b} suggested
$\ell_\nu=2\Delta x$, which leads to $Re=0.4\caln_g^{4/3}$. We note that their value for $\ell_\nu$ is much larger
than the value we inferred from \citet{fede10}, which is in good agreement with the value obtained by \citet{benz08}.

 As a further comparison with results in the literature, we evaluate the wavenumber at which
numerical dissipation begins to affect the results. To make this quantitative,
let $\kd$ be the wavenumber at which $f=1-\delta$.
Equation (\ref{eq:pope}) implies that
\beq
k_{1-\delta}\ell_\nu\simeq 0.47\delta^{1/4}
\label{eq:kod}
\eeq 
to within about 3\% for $\delta< 0.1$. Noting that $k\ell_\nu=(2\pi k'/L)0.6\Delta x$, we find
$\kd'=0.125\delta^{1/4}\caln_g$.
\citet{fede10} concluded that numerical dissipation begins to affect their results at $k'\simeq 40$ in their $1024^3$ simulations. 
Inspection of their results shows that $\delta$ is much less than 0.1 at $k'=40$. Equation (\ref{eq:kod}) implies $\kd'=40$ for $\delta=0.01$,
with only a weak dependence on the value of $\delta$, consistent with their result.

 Finally, we note that for an AMR (adaptive mesh refinement) 
 code like {\sc orion}, the cell size 
 for problems involving self-gravity is generally is set by the requirement that the Jeans length,  $\lj=(\pi\cs^2/G\rho)^{1/2}$, 
 be well resolved \citep{true97}.
 Cells are refined to higher levels if their density exceeds the Truelove-Jeans density, $\rtj$, which is set by the condition
 \beq
 \Delta x=J_{\max}\lj(\rtj),
 \eeq
 where $J_{\max}\leq \frac 14$ is provided by the user, so that
 \begin{equation}
\rtj = \frac{\pi J_{\rm max}^2 c_s^2} { G \Delta x^2}.
\label{eq:rtj}
\end{equation}
 For cases in which the outer scale of the turbulence is set by self-gravity and the Mach number is of order unity, \cite{federrath11b} found that a resolution of 32 zones per Jeans length,
$J_{\max}=1/32$,  is sufficient to see amplification by a turbulent dynamo, whereas a resolution of 16 cells per Jeans length is not. Using a somewhat more dissipative code, \citet{turketal2012} found that a resolution of 64 cells per Jeans length was required. In order to express $\Delta x$ in terms of the local density we have
\beq
\Delta x=J_{\max}\lj(\rho)\left(\frac{\rho}{\rtj}\right)^{1/2}.
\label{eq:dx}
\eeq
At a given level of refinement, the density will range up to $\rtj$. The maximum value of $\Delta x(\rho)$, which
corresponds to a conservative estimate for the viscosity, occurs for $\rho=\rtj$.
Normalizing $J_{\max}$ to 1/64 then gives
 \beq
 \nu_g=1.69\times 10^{24}\left(\frac{J_{\max}}{1/64}\right)^{4/3}\frac{\vtf}{L_\pc^{1/3}}\left(\frac{T_3}{\nh}\right)^{2/3}~~~~~\mbox{cm$^2$ s\e},
 \label{eq:nug1}
 \eeq
 where $\vtf$ is the turbulent velocity in km s\e. This is much larger than the atomic viscosity in equation (\ref{eq:nu}).
 For AMR codes, this equation should be used only if there is at least one level of refinement, so that equation (\ref{eq:dx}) 
 with $\rho=\rtj$ can be used to set
 $\Delta x$; otherwise, equation (\ref{eq:nugrid2}) should be used.
 
 If the outer scale of the turbulence is set by the Jeans length ($L=\lj$), as argued by \citet{federrath11b} for the case of gravitational collapse, the Reynolds number is 
 \beq
 Re=2\left(\frac{\lj}{\Delta x}\right)^{4/3}=512\left(\frac{1/64}{ J_{\max}}\right)^{4/3}
 \label{eq:reg3}
 \eeq
 from equation (\ref{eq:renum}),
 and the viscosity is given by 
 \beq
 \nu_g=2.32\times 10^{23}\vtf\left(\frac{J_{\max}}{1/64}\right)^{4/3}\left(\frac{T_3}{\nh}\right)^{1/2}~~~~~\mbox{cm$^2$ s\e}
 \label{eq:nug2}
 \eeq
from equation (\ref{eq:nug1}).
If the parameters on the right-hand side of this equation are of order unity, this is more than 1000 times larger than the atomic
viscosity; the discrepancy between simulation and reality grows as the density increases.

\subsection{SPH Codes}
\label{app:sphcodes}

We now turn our attention to SPH codes. In SPH codes, the viscosity is determined by the artificial viscosity
that is added to the code.
The standard SPH artificial viscosity
corresponds to a Navier-Stokes viscosity \citep{price2012a,price2012b}
\beq
\nu_\sph=0.1\,\asph\, \cs h_{\rm sm}
\eeq
for subsonic flows, where $\asph$ is the SPH artificial viscosity parameter and
the smoothing length is given in terms of the particle mass, $m_\sph$, as
\beq
h_\sm=h_f (m_\sph/\rho)^{1/3}.
\label{eq:hsm}
\eeq
Here $h_f$ depends on the number of neighbor particles in a kernel, $\caln_{\rm ngb}$, \beq
h_f=\left(\frac{3\caln_{\rm ngb}}{4\pi}\right)^{1/3}\frac{1}{R_{\rm kernel}},
\eeq
where the kernel truncation radius is $R_{\rm kernel}h_\sm$.
\citet{price2012b} adopted $R_{\rm kernel}=2$ and $\caln_{\rm ngb}\simeq 58$ so that $h_f=1.2$, whereas 
Stacy et al (in preparation) adopted $R_{\rm kernel}=1$ and $\caln_{\rm ngb}=200$ so that $h_f=3.63$. 
The parameter $\asph$ can be variable and is often set equal to 0.1 far from shocks,
giving $\nu_\sph=0.01\cs h_{\rm sm}$.
However, \citet{baue12} have argued that this value of $\nu_\sph$ is too low by a factor 6. 
We can resolve this issue by obtaining the value of $\asph$ from equation (\ref{eq:nukh}),
\beq
\asph=4.0\left(\frac{\epsilon^{1/3}}{\cs h_{\rm sm}\kh^{4/3}}\right).
\eeq
We estimate $\kh\simeq 75$ for \cite{price2012b}'s $256^3$ simulation of $\calm=0.3$ turbulence.
He adopted $L=\cs=1$ so that $\epsilon^{1/3}=\calm=0.3$, and the average smoothing length was $h_{\rm sm}=1.2/256$.
Altogether, this gives $\asph=0.8$, slightly larger than the value 0.6 favored by \citet{baue12}, but considerably larger
than 0.1.
Since our estimate is approximate, we shall adopt the value of \citet{baue12},
\beq
\nu_\sph=0.06\, \cs h_{\rm sm}.
\label{eq:nusph}
\eeq
Note that equation (\ref{eq:nusph}) for the SPH viscosity varies linearly with the smoothing length, whereas equation
(\ref{eq:renum}) shows that the grid viscosity varies as $\Delta x^{4/3}$. 
Equation (\ref{eq:hsm}) then gives
\beq
\nu_\sph=1.50\times 10^{23}\left(\frac{h_f m_\sph'^{1/3} T_3^{1/2}}{\nh^{1/3}}\right)~~~~~\mbox{cm$^2$ s\e},
\label{eq:nusph2}
\eeq
where $m_\sph'=m_\sph/(1\,M_\odot)$.
Just as in the case of grid-based viscosity, the numerical viscosity for SPH exceeds the atomic viscosity by more than a factor 1000
if the parameters on the right-hand side are of order unity, and the discrepancy grows as the density increases. 
The adiabatic index of the gas varies from $\gamma\simeq \frac 53$ for gas in the Hubble flow and gas falling into a dark-matter
potential well to $\gamma\simeq 1$ for gas in the protostellar core; here we have set $\gamma=1$ in
our estimate of the SPH viscosity.

For constant density, equation (\ref{eq:nusph}) gives the Reynolds number for SPH,
\beq
Re=17\calm\left(\frac{L}{h_\sm}\right) = 17\calm\left(\frac{\caln_{g,\,\sph}}{h_f}\right),
\eeq
where $\caln_{g,\,\sph}=(\rho L^3/m_\sph)^{1/3}$ is the SPH equivalent to the number of grid cells. The scaling of $Re$ with $\calm$ for SPH codes gives
them an advantage at high Mach numbers \citep{price2012b}.
The fact that $Re$ scales as $\caln_g^{4/3}$ for grid-based codes but only as $\caln_{g,\,\sph}$ for SPH codes means that
grid-based codes become superior to SPH codes at high resolution (V. Springel 2019, private communication). 
For the case in which $\caln_g=\caln_{g,\,\sph}$, the resolution of the grid code must exceed $600\calm^3$ in order for this
advantage to kick in, however.

\subsection{Numerical Resistivity}
\label{app:numres}

The numerical resistivity can be inferred from the values of the numerical viscosity above and of the numerical Prandtl number, $P_m=\nu/\eta$. \citet{les07} found that the numerical Prandtl number for grid-based codes was between 1 and 2, depending on wavenumber.
In their simulations of turbulent amplification of magnetic fields, \citet{federrath11a} inferred that their results were consistent with this conclusion. Subsequently, \citet{federrath11b} studied magnetic field amplification in a gravitationally collapsing cloud. 
They showed that the Jeans length corresponds to the effective outer scale of the turbulence in such a cloud and that the critical magnetic Reynolds number for dynamo action, $\rmcr$, occurred between 16 and 32 cells per Jeans length. 
More generally, \citet{haug04} found $\rmcr=2\pi\times 35P_m^{-1/2}$ for $0.1\la P_m \la 3$. (They defined the magnetic Reynolds number as $R_{m,\,\rm H}=v/(k_f\eta)=vL/(2\pi\eta)$, where $k_f$ is the wavenumber at which the turbulence is forced; this is smaller than the value adopted here by a factor $2\pi$.) Since $Re=R_m/P_m$, we have $Re_{\rm cr}=
2\caln_{g,\,\rm cr}^{4/3}=220/\pmg^{3/2}$. For $\caln_{g,\,\rm cr}$ between 16 and 32, this implies that $\pmg$ is between 1 and 2, just as \citet{les07} found. We shall therefore adopt $\pmg\simeq 1.4$.
Less is known about the magnetic Prandtl number in SPH codes, so we shall adopt $P_m=1.4$ for them also.

\bsp	
\label{lastpage}
\end{document}